\documentclass[12pt]{article}
\usepackage{epsfig}
\usepackage{graphicx}
\usepackage{amssymb}
\usepackage{amsmath}
\usepackage{amsfonts}
\usepackage[dvips]{color}

\newcommand{\R}{\mathbb{R}}
\newcommand{\C}{\mathbb{C}}
\newcommand{\Z}{\mathbb{Z}}

\newcommand{\fa}{\mathfrak{a}}
\newcommand{\fb}{\mathfrak{b}}
\newcommand{\fc}{\mathfrak{c}}

\newcommand{\fz}{\mathfrak{z}}

\newcommand{\fD}{\mathfrak{D}}

\newcommand{\fK}{\mathfrak{K}}

\newcommand{\cB}{\mathcal{B}}

\newcommand{\cP}{\mathcal{P}}

\newcommand{\cR}{\mathcal{R}}
\newcommand{\cS}{\mathcal{S}}
\newcommand{\cT}{\mathcal{T}}

\newcommand{\be}{\begin{equation}}
\newcommand{\ee}{\end{equation}}
\newcommand{\bea}{\begin{eqnarray}}
\newcommand{\eea}{\end{eqnarray}}
\newcommand{\nn}{\nonumber}
\newcommand{\kt}{\rangle}
\newcommand{\br}{\langle}

\newcommand{\ed}{\end{document}}

\newcommand{\rx}{{\rm x}}

\newcommand{\rE}{{\rm E}}

\newcommand{\rT}{M}

\newcommand{\np}{\newpage}

\newcommand{\bi}{\begin{itemize}}
\newcommand{\ei}{\end{itemize}}

\newcommand{\bce}{\begin{center}}
\newcommand{\ece}{\end{center}}

\begin{document}

\title{Spectral Singularities, Biorthonormal Systems,
and a Two-Parameter Family of Complex\\ Point Interactions}

\author{Ali~Mostafazadeh\thanks{E-mail address:
amostafazadeh@ku.edu.tr (corresponding author)} ~and Hossein
Mehri-Dehnavi\thanks{E-mail address: mehrideh@iasbs.ac.ir}
\\
\\
$^*$~Department of Mathematics, Ko\c{c} University, Rumelifeneri
Yolu,\\ 34450 Sariyer, Istanbul, Turkey \\
$ ^\dagger$~Department of Physics, Institute for Advanced Studies in
Basic \\ Sciences, Zanjan 45195-1159, Iran}

\date{ }
\maketitle

\begin{abstract}

A curious feature of complex scattering potentials $v(x)$ is the
appearance of spectral singularities. We offer a quantitative
description of spectral singularities that identifies them with an
obstruction to the existence of a complete biorthonormal system
consisting of the eigenfunctions of the Hamiltonian operator, i.e.,
$-\frac{d^2}{dx^2}+v(x)$, and its adjoint. We establish the
equivalence of this description with the mathematicians' definition
of spectral singularities for the potential
$v(x)=z_-\delta(x+a)+z_+\delta(x-a)$, where $z_\pm$ and $a$ are
respectively complex and real parameters and $\delta(x)$ is the
Dirac delta-function. We offer a through analysis of the spectral
properties of this potential and determine the regions in the space
of the coupling constants $z_\pm$ where it admits bound states and
spectral singularities. In particular, we find an explicit bound on
the size of certain regions in which the Hamiltonian is
quasi-Hermitian and examine the consequences of imposing
$\cP\cT$-symmetry.\vspace{2mm}

\vspace{2mm}

\end{abstract}
\vspace{5mm}

\np

\section{Introduction}

The use of non-Hermitian Hamiltonians in theoretical physics has a
long history. It extends from early attempts to construct
divergence-free relativistic quantum field theories
\cite{indefinite}, to more practical and successful applications in
nuclear and atomic physics \cite{nuclear-phys} and particularly
quantum optics \cite{optics,berry}. During the past ten years there
has been a renewed interest in the study of a special class of
non-Hermitian Hamiltonians that possess a real spectrum. The
best-known examples are the $\cP\cT$-symmetric Hamiltonians
\cite{bender-prl} such as $p^2+ix^3$. These belong to the wider
class of pseudo-Hermitian Hamiltonians $H$ whose adjoint $H^\dagger$
is given by
    \be
    H^\dagger=\eta\,H\eta^{-1},
    \label{ph}
    \ee
for some Hermitian invertible operator $\eta$, \cite{p1,p2-p3}. What
makes pseudo-Hermitian Hamiltonians interesting is that they are
Hermitian with respect to a possibly indefinite inner
product\footnote{This means that $\br \cdot, H\cdot\kt_\eta=\br
H\cdot,\cdot\kt_\eta$.}, namely
    \be
    \br\cdot,\cdot\kt_\eta:=\br\cdot|\eta\cdot\kt,
    \label{inn-ph}
    \ee
where $\br\cdot|\cdot\kt$ is the inner product of the Hilbert space
in which all the relevant operators act. Most of the recent work on
the subject is concentrated on a particular class of
pseudo-Hermitian Hamiltonians, called quasi-Hermitian \cite{quasi},
that satisfy (\ref{ph}) for some positive-definite (metric) operator
$\eta$. In this case, (\ref{inn-ph}) is a positive-definite inner
product, and $H$ becomes Hermitian provided that we define the
physical Hilbert space of the system using the inner product
$\br\cdot,\cdot\kt_\eta$, \cite{jpa-2004b,review}. This allows for
formulating the pseudo-Hermitian representation of quantum mechanics
in which $\cP\cT$-symmetric as well as non-$\cP\cT$-symmetric
quasi-Hermitian Hamiltonians can be employed to describe unitary
quantum systems \cite{review}. The techniques developed in this
framework have so far found interesting applications in relativistic
quantum mechanics \cite{ann}, quantum cosmology \cite{qc}, quantum
field theory \cite{qft}, bound state scattering \cite{matzkin}, and
classical electrodynamics \cite{epl}. But these developments do not
undermine the importance of the requirement that the observables of
a unitary quantum system must be Hermitian with respect to the inner
product of the physical Hilbert space, \cite{review}.

Among the properties of Hermitian operators that make them
indispensable in quantum mechanics is their diagonalizability. For
Hermitian and more generally normal operators (those commuting with
their adjoint), diagonalizability is equivalent to the existence of
an orthonormal basis consisting of the eigenvectors of the operator.
This is more commonly referred to as \emph{completeness}. For a
non-normal (and hence non-Hermitian) operator $H$, diagonalizability
of $H$ means the existence of a basis $\cB^\dagger$ consisting of
(scattering and bound state) eigenfunctions of the adjoint operator
$H^\dagger$ that is biorthonormal to some basis $\cB$ consisting of
the eigenfunctions of $H$, i.e., $\cB$ and $\cB^\dagger$ form a
biorthonormal system of the Hilbert space, \cite{review}. For
brevity we shall call such a biorthonormal system a
\emph{biorthonormal eigensystem} for $H$.

Diagonalizability is a weaker condition than Hermiticity, but
diagonalizable operators with a real and discrete spectrum can be
related to Hermitian operators via similarity transformations
\cite{p2-p3}. This in turn implies that they are quasi-Hermitian
\cite{quasi}. The situation is more complicated when the spectrum is
continuous. A serious difficulty is the emergence of spectral
singularities that conflict with the diagonalizability of the
operator in question \cite{samsonov2}. The aim of this paper is to
elucidate the mechanism by which spectral singularities obstruct the
construction of a biorthonormal eigensystem for the operator. We
shall achieve this aim by obtaining a quantitative measure of lack
of a biorthonormal eigensystem and comparing the latter with the
mathematical condition for the presence of spectral singularities
that is based on the behavior of the Jost functions. In order to
clarify the meaning and consequences of spectral singularities we
shall offer a detailed investigation of the spectral properties of
the Hamiltonian operators of the form
    \be
    H=-\frac{\hbar^2}{2m}\frac{d^2}{d\rx^2}+
    \zeta_+\delta(\rx-\alpha)+\zeta_-\delta(\rx+\alpha),
    \label{H}
    \ee
where $\zeta_\pm$ are complex coupling constants, $\alpha$ is a real
parameter, and $\delta(\rx)$ stands for the Dirac delta-function.

An alternative mechanism that can make a non-Hermitian operator
non-diagonalizable is the emergence of exceptional points. These
correspond to degeneracies where both eigenvalues and eigenvectors
coalesce \cite{ep}. Exceptional points have various physical
implications \cite{optics,berry,ep2}. But they must not be confused
with spectral singularities. Unlike exceptional points that can be
present for operators with a discrete spectrum (in particular matrix
Hamiltonians), spectral singularities are exclusive features of
certain operators having a continuous part in their spectrum. As we
will see in Sections 2 and 3, for an operator having a spectral
singularity we can still define two linearly independent
(scattering) eigenfunctions for each eigenvalue, nevertheless it is
impossible to construct a biorthonormal eigensystem for the
operator. To the best of our knowledge, physical meaning of spectral
singularities and their possible applications have not been
previously studied. This is the subject of Ref.~\cite{p89} where the
results of the present paper have been used to develop a physical
interpretation for spectral singularities.

Ref.~\cite{samsonov} uses the mathematical theory of spectral
singularities developed in \cite{naimark-review,ljance} to emphasize
their relevance to the recent attempts at using complex scattering
potentials to define unitary quantum systems. The results of
\cite{samsonov} are, however, confined to potentials defined on the
half-line $\rx\geq 0$, where the Hamiltonian operator acts in the
Hilbert space of square-integrable functions $\psi: [0,\infty)\to\C$
satisfying the boundary condition $\psi(0)=0$. Furthermore, due to
the nature of the concrete potentials studied in \cite{samsonov}, it
has not been possible to construct bases of the corresponding
Hamiltonian and its adjoint and show by explicit calculation how the
presence of a spectral singularity obstructs the existence of a
biorthonormal eigensystem. This is quite essential, because for the
cases that the spectrum is real, the availability of a biorthonormal
eigensystem is a necessary condition for the existence of an
associated metric operator and the quasi-Hermiticity of the
Hamiltonian, \cite{review}.

The only thoroughly studied example of a complex scattering
potential that is defined on the whole real line and can lead to
spectral singularities is the single-delta-function potential with a
complex coupling \cite{jpa-2006b}. The Hamiltonian operator is given
by (\ref{H}) with $\alpha=\zeta_-=0$. It develops a spectral
singularity if and only if the coupling constant ($\zeta_+$) is
imaginary. In particular, for the cases that the real part of
$\zeta_+$ is positive, the bound states are also lacking and the
Hamiltonian is quasi-Hermitian. The complex single-delta-function
potentials provide a class of manifestly non-$\cP\cT$-symmetric
Hamiltonians with a continuous spectrum that happen to be
quasi-Hermitian. An advantage of considering complex
double-delta-function potentials is that their space of coupling
constants has a subspace, given by $\zeta_+=\zeta_-^*$, where the
Hamiltonian is $\cP\cT$-invariant. Therefore, these potentials
provide an opportunity to investigate the significance of
$\cP\cT$-symmetry \cite{bender-review}.

The spectral properties of the $\cP\cT$-symmetric double- and
multiple-delta-function potentials have been studied in
\cite{jones,ahmed,albaverio,demiralp}. The results are, however,
confined to the determination of the (scattering and bound-state)
spectrum of these potentials, and no attempt has been made to decide
if these potentials lead to spectral singularities.

In the present paper we will try to obtain a map of the space
$\C^2=\R^4$ of the coupling constants $\zeta_\pm$ that specifies the
regions corresponding to the existence of bound states and spectral
singularities. We will in particular investigate the intersection of
these regions with the two-dimensional $\cP\cT$-symmetric subspace:
$\zeta_-=\zeta_+^*$. The following is an outline of the results we
report in this paper. In Section~2, we obtain an explicit
quantitative measure of the existence of biorthonormal eigensystems
and compare the latter with the known condition of the presence of
spectral singularities. Here we also provide a useful
characterization of spectral singularities and bound states for
complex scattering potentials. Section~3 treats the spectral
properties of the double-delta-function potentials. It consists of
four subsections in which we obtain the regions in the space of
coupling constants where spectral singularities and bound states
exist, find their location in the spectrum of the operator, and
determine a lower bound on the size of certain regions in $\C^2$
where the operator (\ref{H}) is quasi-Hermitian. Section~4 presents
our concluding remarks.

\section{Spectral Singularities}

Consider a complex-valued potential $v:\R\to\C$ depending on a set
of complex coupling constants $z_1,z_2,\cdots,z_d$ such that $v^*$
is obtained by complex-conjugating the coupling constants in the
expression for $v$. Suppose that $v$ decays rapidly\footnote{For the
purpose of the present paper, we assume that as $|x|\to\infty$ we
have $|v(x)|\leq \exp[-\epsilon\sqrt{|x|}]$ for some
$\epsilon\in\R^+$. As far as the general properties related with
spectral singularities, all the results hold for the less rapidly
decaying potentials that satisfy $\int_{-\infty}^\infty dx\:
(1+|x|)|v(x)|<\infty$. See \cite{TB-1999,guseinov}.} as
$|x|\to\infty$ and that the spectrum of the corresponding
Hamiltonian operator,
    \be
    H=-\frac{d^2}{dx^2}+v(x),~~~~~~~~~x\in\R,
    \label{second-order}
    \ee
is the set of nonnegative real numbers.\footnote{We shall consider
the more general case that the spectrum involves eigenvalues (with
square-integrable eigenfunctions) at the end of this section.} Let
$\psi^{\vec z}_{\fa k}(x)$ denote the (generalized) eigenfunctions
of $H$, i.e., linearly-independent bounded solutions of
    \be
    H\psi^{\vec z}_{\fa k}(x)=k^2\psi^{\vec z}_{\fa k}(x),
    \label{eg-va-zero}
    \ee
where $k\in\R^+$ and $\fa\in\{1,2\}$ are respectively the spectral
and degeneracy\footnote{The spectrum is necessarily
doubly-degenerate.} labels and $\vec z:=(z_1,z_2,\cdots,z_d)$.

By definition, $H$ is diagonalizable, if $\psi^{\vec z}_{\fa k}(x)$
together with a set of (generalized) eigenfunctions $\phi^{\vec
z}_{\fa k}(x)$ of $H^\dagger$ form a complete biorthonormal system
$\{\psi^{\vec z}_{\fa k},\phi^{\vec z}_{\fa k}\}$, i.e., they
satisfy
    \be
    \br\phi^{\vec z}_{\fa k}|\psi^{\vec z}_{\fb
    q}\kt=\delta_{\fa\fb}\:\delta(k-q),~~~~~~~
    \sum_{\fa=1}^2\int_0^\infty dk\;|\psi^{\vec z}_{\fa k}\kt\br\phi^{\vec z}_{\fa
    k}|=1,
    \label{bi-ortho}
    \ee
where $\br\cdot|\cdot\kt$ is the usual $L^2$-inner product. The
biorthonormality relations (\ref{bi-ortho}) imply the spectral
representation of $H$,
    \be
    H=\sum_{\fa=1}^2\int_0^\infty dk\;k^2|\psi^{\vec z}_{\fa k}\kt\br\phi^{\vec z}_{\fa
    k}|,
    \label{sp-rep}
    \ee
as well as the eigenfunction expansion:
    \be
    f(x)=\sum_{\fa=1}^2\int_0^\infty dk\;f_{\fa k}\:\psi^{\vec z}_{\fa
    k}(x),
    \label{eg-fn-exp}
    \ee
where $f:\R\to\C$ is a test function and
    \be
    f_{\fa k}:=\br\phi^{\vec z}_{\fa k}|f\kt.
    \label{coeff}
    \ee

Because $H^\dagger=-\frac{d^2}{dx^2}+v(x)^*$, $\psi^{\vec z^*}_{\fa
k}$ are the eigenfunctions of $H^\dagger$. This in turn means that
$\phi^{\vec z}_{\fa k}$ must be a linear combination of $\psi^{\vec
z^*}_{\fa k}$, i.e., there are $J_{\fa\fb}\in\C$ satisfying
    \be
    \phi^{\vec z}_{\fa k}=\sum_{\fb=1}^2 J_{\fa\fb}\,\psi^{\vec z^*}_{\fb
    k}.
    \label{phi-psi}
    \ee
In view of (\ref{bi-ortho}), there must exist $K_{\fa\fb}\in\C$ such
that
    \be
    \br \psi^{\vec z^*}_{\fa k}|\psi^{\vec z}_{\fb
    q}\kt=K_{\fa\fb}\,\delta(k-q).
    \label{psi-psi-2}
    \ee
Furthermore, if we respectively denote by $I$, $J$ and $K$ the
$2\times 2$ identity matrix and the matrices with entries
$J_{\fa\fb}$ and $K_{\fa\fb}$, we find $J^*K=I$. In particular, $K$
must be an invertible matrix and $J_{\fa\fb}=K^{-1*}_{\fa\fb}$. We
can write this relation in the form
    \be
    J_{\fa\fb}=\frac{\tilde K_{\fa\fb}^*}{\det(K)^*},
    \label{JM}
    \ee
where $\tilde K$ is the transpose of the matrix of cofactors of $K$.
It satisfies
    \[\left(\begin{array}{cc}
    \br \psi^{\vec z^*}_{2 k}|\psi^{\vec z}_{2 q}\kt &
    -\br \psi^{\vec z^*}_{1 k}|\psi^{\vec z}_{2 q}\kt\\
    -\br \psi^{\vec z^*}_{2 k}|\psi^{\vec z}_{1 q}\kt &
    \br \psi^{\vec z^*}_{1 k}|\psi^{\vec z}_{1
    q}\kt\end{array}\right)=\tilde K\:\delta(k-q).\]

We can use (\ref{JM}) and (\ref{phi-psi}) to express (\ref{coeff})
as
    \be
    f_{\fa k}=\frac{1}{\det(K)}\sum_{\fb=1}^2\tilde K_{\fa\fb}\br
    \psi^{\vec z^*}_{\fb k}|f\kt.
    \label{coeff-2}
    \ee
According to this equation if $\det(K)=0$, the eigenfunction
expansion (\ref{eg-fn-exp}) breaks down; the eigenfunctions
$\psi^{\vec z}_{\fa k}$ do not form a complete set; and $H$ is not
diagonalizable. We identify this situation with the presence of
spectral singularities:
    \be
    \mbox{\begin{tabular}{|c|}
    \hline
    ~\emph{Spectral singularities are points $k^2$ of the continuous
    spectrum of $H$ where $\det(K)=0$}.\\
    \hline
    \end{tabular}}
    \label{def-1}
    \ee

Because of (complex) analyticity property of the eigenfunctions
$\psi^{\vec z}_{\fa k}$, $\det(K)$ is an analytic function of $k$.
Therefore the (real) zeros of $\det(K)$ are isolated points forming
a countable subset of the real line. Moreover, because $v$ is a
bounded function decaying rapidly away from zero, the eigenfunctions
tend to plane waves as $k$ becomes large. This shows that $\det(K)$
does not have arbitrarily large zeros (for fixed $\vec z$). As a
result, the zeros of $\det(K)$ are not only isolated but actually
finite in number. In other words, depending on the values of the
coupling constants $z_1,z_2,\cdots,z_d$, $\det(K)$ may have no
(real) zeros in which case spectral singularities do not arise and
$H$ is diagonalizable, or a finite number of (non-vanishing real)
zeros $\kappa_1,\kappa_2,\cdots,\kappa_\mu$ in which case
$\kappa_1^2,\kappa_2^2,\cdots,\kappa_\mu^2$ are spectral
singularities and $H$ is not diagonalizable.

In general the space of coupling constants can be divided into two
regions, namely the \emph{singular region} where $H$ has spectral
singularities and the \emph{regular region} where it is
diagonalizable.

In mathematics literature a spectral singularity is defined as
follows:
    \begin{itemize}
    \item[] \textbf{Definition~1}: \emph{An element $E_\star$ of the
    (continuous) spectrum of $H$ is called a spectral singularity
    if the integral kernel of the resolvent operator:
    $(H-E)^{-1}$, i.e., the Green's function $\br x|(H-E)^{-1}|y\kt$, is
    an unbounded function in every small open neighborhood of $E_\star$,
    but $E_\star$ is not an eigenvalue of $H$ with a square-integrable
    eigenfunction }\cite{guseinov}.\footnote{Therefore spectral singularities are certain
    poles of $\br x|(H-E)^{-1}|y\kt$.}
    \end{itemize}
There is a rather general theory of spectral singularities for the
differential operators of the form (\ref{second-order}) where the
spectral singularities are characterized as the real zeros of
certain analytic functions
\cite{naimark-book2,naimark-review,kemp,schwartz,ljance,TB-1999}.
For the case that the operator acts in $L^2(\R)$, this is the
Wronskian,
    \be
    W[\psi_{k-},\psi_{k+}]:=
    \psi_{k-}(x)\psi_{k+}'(x)-\psi_{k-}'(x)\psi_{k+}(x)=
    \psi_{k-}(0)\psi_{k+}'(0)-\psi_{k-}'(0)\psi_{k+}(0),~~~~~~
    \label{wronskian}
    \ee
of the Jost solutions $\psi_{k\pm}$ of the eigenvalue equation
$H\psi=k^2\psi$. These are defined in terms of their asymptotic
behavior
    \be
    \psi_{k-}(x)\to e^{-ikx}~~{\rm
    for}~~x\to-\infty,\quad\quad\quad\quad
    \psi_{k+}(x)\to e^{ikx}~~{\rm for}~~x\to\infty.
    \label{jost}
    \ee
More specifically, we have \cite{guseinov}
    \be
    \mbox{\begin{tabular}{|c|}
    \hline
    ~\emph{Spectral singularities are the real
    (non-vanishing) zeros of $W[\psi_{k-},\psi_{k+}]$}.\\
    \hline
    \end{tabular}}
    \label{def-2}
    \ee
This description of spectral singularities seems to differ from the
one given in (\ref{def-1}). In Section~3, we demonstrate the
equivalence of the two descriptions for the double-delta-function
potential by explicit calculations. The following calculation shows
how the description (\ref{def-1}) relates to Definition~1. Using
(\ref{bi-ortho}), (\ref{sp-rep}), (\ref{phi-psi}) and (\ref{JM}), we
have
    \bea
    \br x|(H-E)^{-1}|y\kt&=&\sum_{\fa=1}^2\int_0^\infty dk\;
    \frac{\psi^{\vec z}_{\fa k}(x)\phi^{\vec z}_{\fa k}(y)^*}{k^2-E}=
    \sum_{\fa,\fb=1}^2\int_0^\infty dk\;
    \frac{J_{\fa\fb}^*\psi^{\vec z}_{\fa k}(x)
    \psi^{\vec z^*}_{\fb k}(y)^*}{k^2-E}\nn\\
    &=&
    \sum_{\fa,\fb=1}^2\int_0^\infty dk\;
    \frac{\tilde K_{\fa\fb}\psi^{\vec z}_{\fa k}(x)
    \psi^{\vec z^*}_{\fb k}(y)^*}{\det(K)(k^2-E)}.\nn
    \eea

In the remainder of this section we provide a useful
characterization of the spectral singularities and bound states.

Because $|v(x)|$ decays rapidly as $|x|\to\infty$, solutions of
(\ref{eg-va-zero}) have the asymptotic behavior:
    \be
    \psi_{k\fa}^{\vec z}(x)\to A_\pm e^{ikx}+B_\pm e^{-ikx}~~~~~
    {\rm for}~~~~x\to\pm\infty,
    \label{gen-asym}
    \ee
where $A_\pm$ and $B_\pm$ are possibly $k$-dependent complex
coefficients. If we denote the coefficients $A_\pm$ and $B_\pm$ for
the Jost solutions $\psi_{k\pm}$ by $A_\pm^{\pm}$ and $B_\pm^\pm$,
we can express (\ref{jost}) as
    \be
    A_+^+=B_-^-=1,~~~~~A_-^-=B_+^+=0.
    \label{jost2}
    \ee

Next, we let $\rT=({\rT}_{\fa\fb})$ be the possibly $k$-dependent
$2\times 2$ transfer matrix \cite{razavy} satisfying
    \be
    \left(\begin{array}{c}
    A_+\\
    B_+\end{array}\right)=\rT\left(\begin{array}{c}
    A_-\\
    B_-\end{array}\right),
    \label{S-matrix}
    \ee
and use this relation and Eqs.~(\ref{gen-asym}) and (\ref{jost2}) to
obtain
    \be
    A_-^+=\frac{{\rT}_{22}}{\det \rT},~~~~B_-^+=-\frac{{\rT}_{21}}{\det \rT},
    ~~~~A_+^-={\rT}_{12},~~~~B_+^-={\rT}_{22}.
    \label{S-mat-3}
    \ee
Inserting these equations in (\ref{gen-asym}), we find
    \bea
    \psi_{k-}(x)&\to&
    {\rT}_{12}\,e^{ikx}+{\rT}_{22}\,e^{-ikx}~~~~~
    {\rm for}~~~~~x\to\infty,
    \label{jost3-}\\
    \psi_{k+}(x)&\to&
    \frac{{\rT}_{22}\,e^{ikx}-{\rT}_{21}\,e^{-ikx}}{\det \rT}~~~~~
    {\rm for}~~~~~x\to-\infty,
    \label{jost3}
    \eea
Because according to Abel's theorem \cite{boyce}, the Wronskian of
solutions of (\ref{eg-va-zero}) is independent of $x$, we can use
the asymptotic formulas for the Jost solutions to compute their
Wronskian. We use Eqs.~(\ref{jost}), (\ref{jost3-}) and
(\ref{jost3}), to perform this calculation first for $x\to\infty$
and then for $x\to-\infty$. This gives
    \bea
    W[\psi_{k-},\psi_{k+}]&=& 2ik\,{\rT}_{22}(k),
    \label{wronskian2}\\
    W[\psi_{k-},\psi_{k+}]&=&\frac{2ik\,{\rT}_{22}(k)}{\det
    \rT(k)},
    \label{wronskian2-}
    \eea
where we have made the $k$-dependence of ${\rT}_{22}$ and $\rT$
explicit. A direct consequence of (\ref{wronskian2}) and
(\ref{wronskian2-}) is
    \be
    \det \rT(k)=1.
    \label{det-S=}
    \ee
More importantly, we have the following characterization of spectral
singularities that follows from (\ref{def-2}) and
(\ref{wronskian2}).
    \be
    \mbox{
    \begin{tabular}{|c|}
    \hline
    ~\emph{Spectral singularities are given by $k^2$ where $k$ is a
    (non-vanishing) real zero of ${\rT}_{22}(k)$}.\\
    \hline
    \end{tabular}}
    \label{desc-ss}
    \ee

Finally, consider the more general case that the Hamiltonian
operator (\ref{second-order}) has, in addition to a continuous
spectrum corresponding to $k\in\R^+$, a possibly complex discrete
spectrum. The latter corresponds to the square-integrable solutions
of (\ref{eg-va-zero}) that represent bound states. It is not
difficult to show that the spectral label corresponding to these
bound states are also zeros of ${\rT}_{22}(k)$, but unlike the zeros
associated with the spectral singularities these must have a
positive imaginary part. In other words, we have the following
characterization of the bound states.
    \be
    \mbox{
    \begin{tabular}{|c|}
    \hline
    ~\emph{Bound state energies are
    given by $k^2$ where $k$ is a zero of ${\rT}_{22}(k)$ with
    $\Im(k)>0$}.\\
    \hline
    \end{tabular}}
    \label{desc-bs}
    \ee

\section{The Double-Delta-Function Potential}

\subsection{Eigenfunctions for Scattering and Bound States}

Consider the time-independent Schr\"odinger equation
    \be
    \left[-\frac{\hbar^2}{2m}\frac{d^2}{d\rx^2}+
    \zeta_+\delta(\rx-\alpha)+\zeta_-\delta(\rx+\alpha)\right]\psi=
    \rE\psi.
    \label{eg-va}
    \ee
Let $\ell$ be an arbitrary length scale and introduce the
dimensionless quantities
    \be
    z_\pm:=\frac{2m\ell\zeta_\pm}{\hbar^2},~~~x:=\frac{\rx}{\ell},~~~
    a:=\frac{\alpha}{\ell},
    ~~~E:=\frac{2m\ell^2\rE}{\hbar^2}.
    \ee
Then (\ref{eg-va}) takes the form
    \be
    -\psi''+[z_+\delta(x-a)+z_-\delta(x+a)]\psi=E\psi.
    \label{eg-va-2}
    \ee

We can write solutions of (\ref{eg-va-2}) as
    \bea
    \psi(x)&=&\left\{\begin{array}{ccc}
    \psi^-(x)&{\rm for}& x<-a,\\
    \psi^0(x)&{\rm for}& |x|\leq a,\\
    \psi^+(x)&{\rm for}& x>a,\end{array}\right.
    \label{wf1}\\
    \psi^\nu(x)&=&A_\nu e^{ikx}+B_\nu
    e^{-ikx},~~~~~~\nu\in\{-,0,+\},
    \label{wf2}
    \eea
where $k:=\sqrt{E}$ and without of loss of generality we require
that the principal argument of $k$ belongs to $[0,\pi)$.

To determine the matching conditions at $x=\pm a$, we demand that
$\psi$ be continuous, i.e.,
    \be
    \psi^-(-a)=\psi^0(-a),~~~~~
    \psi^0(a)=\psi^+(a).
    \label{conti}
    \ee
Furthermore, we integrate both sides of (\ref{eg-va-2}) over the
intervals $[\mp a-\epsilon,\mp a+\epsilon]$ and take the limit
$\epsilon\to 0$ in the resulting formulas to find
    \be
    {\psi^-}'(-a)-{\psi^0}'(-a)+z_-\psi^0(-a)=0,~~~~~
    {\psi^0}'(a)-{\psi^+}'(a)+z_+\psi^0(a)=0.
    \label{der}
    \ee
Introducing
    \be
    w_\pm:=\frac{i z_\pm}{2k},
    \label{w=}
    \ee
and inserting (\ref{wf1}) and (\ref{wf2}) in (\ref{conti}) and
(\ref{der}) yield the desired matching conditions that we can write
in the form
    {\small\be
    \left(\begin{array}{c}
    A_-\\
    B_-\end{array}\right)=\left(\begin{array}{cc}
    1+w_-& w_-e^{2iak}\\
    -w_-e^{-2iak} & 1-w_-\end{array}\right)\left(\begin{array}{c}
    A_0\\
    B_0\end{array}\right),~~~
    \left(\begin{array}{c}
    A_+\\
    B_+\end{array}\right)=\left(\begin{array}{cc}
    1-w_+& -w_+e^{-2iak}\\
    w_+e^{2iak} & 1+w_+\end{array}\right)\left(\begin{array}{c}
    A_0\\
    B_0\end{array}\right).
    \label{match}
    \ee}
In light of these relations, the matrix $\rT$ satisfying
(\ref{S-matrix}) reads
    \be
    \rT=
    \left(\begin{array}{cc}
    1-w_--w_++(1-e^{-4iak})w_-w_+ &
    2iw_-w_+\sin(2ak)-w_-e^{2iak}-w_+e^{-2iak}\\
    -2iw_-w_+\sin(2ak)+w_-e^{-2iak}+w_+e^{2iak} &
    1+w_-+w_++(1-e^{4iak})w_-w_+\end{array}\right).
    \label{S}
    \ee
It is easy to check that indeed $\det(\rT)=1$.

Next, we let $\vec z$ stand for $(z_-,z_+)$ and use $\psi^{\vec
z}_{1k}$ and $\psi_{2k}^{\vec z}$ to denote the eigenfunctions
obtained by setting $A_0=(2\pi)^{-1/2}, B_0=0$ and
$A_0=0,B_0=(2\pi)^{-1/2}$ respectively. Then,
    {
    \bea
    \psi_{1k}^{\vec z}(x)&=&(2\pi)^{-1/2}\times\left\{
    \begin{array}{ccc}
    (1+w_-)e^{ikx}-w_-e^{-ik(x+2a)}&{\rm for}& x<-a,\\
    e^{ikx}&{\rm for}& |x|\leq a,\\
    (1-w_+)e^{ikx}+w_+e^{-ik(x-2a)}&{\rm for}& x>a,\end{array}\right.
    \label{psi1}\\  && \nn \\
    \psi_{2k}^{\vec z}(x)&=&(2\pi)^{-1/2}\times\left\{
    \begin{array}{ccc}
    (1-w_-)e^{-ikx}+w_-e^{ik(x+2a)}&{\rm for}& x<-a,\\
    e^{-ikx}&{\rm for}& |x|\leq a,\\
    (1+w_+)e^{-ikx}-w_+e^{ik(x-2a)}&{\rm for}& x>a.\end{array}\right.
    \label{psi2}
    \eea}%
We can construct a set of eigenfunctions of $H^\dagger$ by taking
$z_\pm$ to $z_\pm^*$ or $w_\pm$ to $-w_\pm^*$ in these relations.
They are given by
    {
    \bea
    \psi_{1k}^{\vec z^*}(x)&=&(2\pi)^{-1/2}\times\left\{
    \begin{array}{ccc}
    (1-w_-^*)e^{ikx}+w_-^*e^{-ik(x+2a)}&{\rm for}& x<-a,\\
    e^{ikx}&{\rm for}& |x|\leq a,\\
    (1+w_+^*)e^{ikx}-w_+^*e^{-ik(x-2a)}&{\rm for}& x>a,\end{array}\right.
    \label{phi1}\\  && \nn \\
    \psi_{2k}^{\vec z^*}(x)&=&(2\pi)^{-1/2}\times\left\{
    \begin{array}{ccc}
    (1+w_-^*)e^{-ikx}-w_-^*e^{ik(x+2a)}&{\rm for}& x<-a,\\
    e^{-ikx}&{\rm for}& |x|\leq a,\\
    (1-w_+^*)e^{-ikx}+w_+^*e^{ik(x-2a)}&{\rm for}& x>a.\end{array}\right.
    \label{phi2}
    \eea}%

\subsection{Characterization of Spectral Singularities}

In this subsection we use (\ref{def-1}) to determine the spectral
singularities of the double-delta-function potential. This requires
computes $\br\psi^{\vec z^*}_{\fa,k}|\psi^{\vec z}_{\fb,q}\kt$ for
all $\fa,\fb\in\{1,2\}$. Using (\ref{psi1}) and (\ref{psi2}) and the
identities
    \[\int_\nu^\infty e^{i\mu x}dx=\pi\delta(\mu)+\frac{i
    e^{i\mu\nu}}{\mu},~~~\int_{-\infty}^\nu e^{i\mu x}dx=\pi\delta(\mu)-\frac{i
    e^{i\mu\nu}}{\mu},\]
we find
    \be
    \left(\begin{array}{cc}
    \br\psi^{\vec z^*}_{1k}|\psi^{\vec z}_{1q}\kt &
    \br\psi^{\vec z^*}_{1k}|\psi^{\vec z}_{2q}\kt\\
    \br\psi^{\vec z^*}_{2k}|\psi^{\vec z}_{1q}\kt &
    \br\psi^{\vec z^*}_{2k}|\psi^{\vec z}_{2q}\kt\end{array}\right)=\delta(k-q)K,
    \label{phi-psi-2}
    \ee
where $K=(K_{ij})$ is a $2\times 2$ matrix with entries
    \bea
    K_{11}&=&K_{22}=1-w_-^2-w_+^2=1+\frac{z_-^2+z_+^2}{4k^2},
    \label{e-K11}\\
    K_{12}&=&w_-(1-w_-)e^{2iak}-w_+(1+w_+)e^{-2iak}\nn\\
    &=&(4k^2)^{-1}\left[
    iz_-(2k-iz_-)e^{2iak}-iz_+(2k+iz_+)e^{-2iak}\right],
    \label{e-K12}\\
    K_{21}&=&-w_-(1+w_-)e^{-2iak}+w_+(1-w_+)e^{2iak}\nn\\
    &=&(4k^2)^{-1}\left[
    -iz_-(2k+iz_-)e^{-2iak}+iz_+(2k-iz_+)e^{2iak}\right].
    \label{e-K21}
    \eea
In the $\cP\cT$-symmetric case, where $z_+=z_-^*=:z\neq 0$, $K$ is a
real matrix, and
    \bea
    K_{11}&=&K_{22}=1+\frac{\Re(z^2)}{2k^2},
    \label{k11-pt}\\
    K_{12}&=&(2k^2)^{-1}\Im\left[z(2k+iz)e^{-2iak}\right],
    \label{k12-pt}\\
    K_{21}&=&
    (2k^2)^{-1}\Im\left[z(-2k+iz)e^{2iak}\right].
    \label{k21-pt}
    \eea

The fact that $K$ is not generally diagonal shows that $\{\psi^{\vec
z}_{\fa k},\psi^{\vec z^*}_{\fb q}\}$ is not a biorthonormal system.
To construct the basis biorthonormal to $\{\psi^{\vec z}_{\fa k}\}$
we transform $\psi^{\vec z^*}_{\fa k}$ according to
    $$\psi^{\vec z^*}_{\fa k}\to \phi^{\vec z}_{\fa k}:=\sum_{\fb=1}^2
    J_{\fa \fb}\,\psi^{\vec z^*}_{\fb k},$$
and fix the coefficients $J_{\fa \fb}$ by demanding that
$\{\psi^{\vec z}_{\fa k},\phi_{\fa k}^{\vec z}\}$ be a biorthonormal
system. As we explained in Section~2, in terms of $K$ this condition
takes the form $\delta_{\fa\fb}=\sum_{\fc=1}^2
J_{\fa\fc}^*K_{\fc\fb}$. Therefore, a basis biorthonormal to
$\{\psi^{\vec z}_{\fa k}\}$ exists provided that the matrix $K$ is
invertible, and the matrix $J$ of coefficients $J_{\fa\fb}$ has the
form $J=K^{-1*}$.

The nonzero real values of $k$ for which $K$ is a singular matrix
give the spectral singularities of $H$. These are the non-vanishing
real zeros of $\det(K)$ that we can obtain using (\ref{e-K11}) --
(\ref{e-K21}):
    \be
    \det(K)=1+\frac{z_-^2+z_+^2}{4k^2}+
    \frac{z_-^2z_+^2}{8k^4}+\frac{z_-z_+}{2k^2}\left[
    \left(1-\frac{z_-z_+}{4k^2}\right)\cos(4ak)+
    \left(\frac{z_-+z_+}{2k}\right)\sin(4ak)\right]=0.
    \label{det-K}
    \ee
If either $z_-=0$ and $z:=z_+$ or $z_+=0$ and $z:=z_-$, this
equation reduces to
    \[ 1+\frac{z^2}{4k^2}=0.\]
Therefore, for pure imaginary $z$ there is a spectral singularity
located at $k=\pm iz/2=|z|/2$. This agrees with the results for the
single delta-function potential \cite{jpa-2006b}.

For the $\cP\cT$-symmetric case ($z_+=z_-^*=:z$), we have
    \be
    \det(K)=1+\frac{\Re(z^2)}{2k^2}+
    \frac{|z|^4}{8k^4}+\frac{|z|^2}{2k^2}\left[
    \left(1-\frac{|z|^2}{4k^2}\right)\cos(4ak)+
    \left(\frac{\Re(z)}{k}\right)\sin(4ak)\right]=0.
    \label{det-K-PT}
    \ee
In particular if $z$ is purely imaginary, i.e., $z=i\sigma$ for some
$\sigma\in\R$,
    \be
    \det(K)=\cos^2(2ak)+\Big(1-\frac{\sigma^2}{2k^2}\Big)^2\sin^2(2ak).
    \label{det-K-PT-imaginary}
    \ee
Therefore, $\det(K)=0$ iff $\cos(2ak)=0$ and $k=|\sigma|/\sqrt 2$.
This implies that there is a spectral singularity for
$k=|\sigma|/\sqrt 2=|z|/\sqrt 2$ iff $\sigma$ takes one of the
following values
    \be
    \sigma_n:=\frac{\pi(2n+1)}{2\sqrt 2~a},~~~~~n\in\Z.
    \label{sigma-n}
    \ee
In summary, for the case that $z_+=-z_-=:z$ is purely imaginary, $H$
has a single spectral singularity given by
    \be
    E_\star=\frac{\sigma_n^2}{2}=\left[\frac{(2n+1)\pi}{4a}\right]^2,
    \label{ss-imaginary-pt}
    \ee
if $z=i\sigma_n$ for some $n\in\Z$. Otherwise it does not have any
spectral singularities.

Next, consider the general $\cP\cT$-symmetric case where
$z_+=z_-^*=:z$, and $z$ need not be purely imaginary. In this more
general case, we rewrite (\ref{det-K-PT}) in the form
    \be
    \det(K)=|f(z,a,k)|^2,
    \label{det-K-PT2}
    \ee
where
    \be
    f(z,a,k):=\frac{|z|^2\sin(2ak)}{2k^2}+e^{-2iak}\left(\frac{\Re(z)}{k}-i\right).
    \ee
It is easy to compute
    \bea
    \Re[f(z,a,k)]&=&\left(\frac{|z|^2}{2k^2}-1\right)\sin(2ak)+\frac{\Re(z)}{k}\,\cos(2ak),
    \label{real}\\
    \Im[f(z,a,k)]&=&-\left[\cos(2ak)+\frac{\Re(z)}{k}\,\sin(2ak)\right].
    \label{imaginary}
    \eea

In view of (\ref{det-K-PT2}), $\det(K)=0$ iff
$\Re[f(z,a,k)]=\Im[f(z,a,k)]=0$. Imposing $\Im[f(z,a,k)]=0$, we have
    \be
    \cos(2ak)=-\frac{\Re(z)}{k}\,\sin(2ak),
    \label{q1}
    \ee
which in particular implies $\sin(2ak)\neq 0$. Moreover,
$\cos(2ak)=0$ iff $\Re(z)=0$. In light of $\sin(2ak)\neq 0$ and
(\ref{q1}), $\Re[f(z,a,k)]=0$ gives
    \be
    -\Re(z)^2+\Im(z)^2=2k^2.
    \label{q2}
    \ee
This implies that \emph{if $|\Re(z)|\geq|\Im(z)|$, there is no
spectral singularity.}

\begin{figure}[t]
\begin{center} 
\includegraphics[scale=.75,clip]{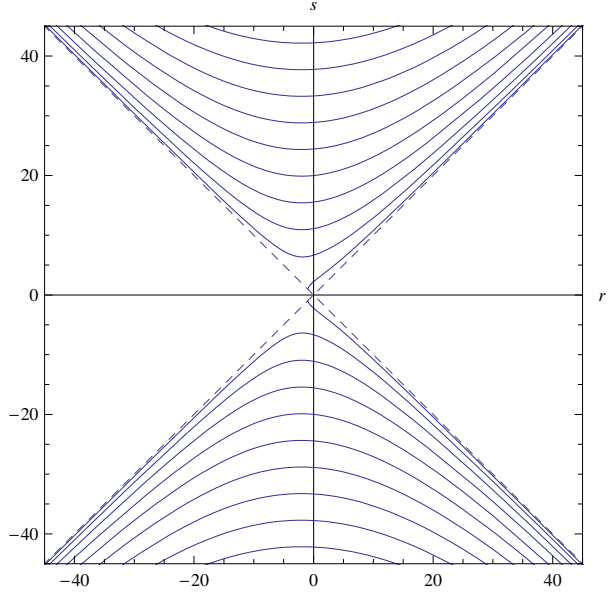}
\parbox{14cm}{\caption{Curves in the $r$-$s$ plane giving the location
of the spectral singularities for the general $\cP\cT$-symmetric
double-delta-function potential. The dashed lines are the asymptotes
$s=\pm r$. The intersection of the curves with the $s$-axis
corresponds to the spectral singularities given by
Eq.~(\ref{ss-imaginary-pt}). \label{fig1}}}\end{center}
\end{figure}

Next, we solve (\ref{q2}) for $k$ to obtain
    \be
    k=\sqrt{\frac{-\Re(z)^2+\Im(z)^2}{2}},
    \label{q5}
    \ee
and express (\ref{q1}) as
    \be
    \Re(z)=-k\cot(2ak).
    \label{q3}
    \ee
Inserting (\ref{q5}) in (\ref{q3}) yields a necessary and sufficient
condition for the existence of a spectral singularity, namely
    \be
    2\Re(z)\tan\left(a\sqrt{2[-\Re(z)^2+\Im(z)^2]}\right)+
    \sqrt{2[-\Re(z)^2+\Im(z)^2]}=0.
    \label{q6}
    \ee
Introducing the variables
    \be
    r:=2a\Re(z),~~~~~~~~~s:=2a\Im(z),~~~~~~~~~
    t:=a\sqrt{2[-\Re(z)^2+\Im(z)^2]},
    \label{r-s}
    \ee
we can express (\ref{q6}) in the form
    \be
    r=-t\cot t,
    \label{r=}
    \ee
and establish
    \be
    s=\pm t\sqrt{1+\csc^2t}.
    \label{s=}
    \ee
Figure~\ref{fig1} shows a plot of the parametric curve defined by
(\ref{r=}) and (\ref{s=}). It consists of an infinite set of
disjoint open curves with asymptotes $s=\pm r$ in the $r$-$s$ plane.
The points on these curves correspond to the values of the coupling
constant $z$ for which a spectral singularity appears.

Next, consider the general not necessarily $\cP\cT$-symmetric case.
Generalizing our treatment of the $\cP\cT$-symmetric case, we use
(\ref{det-K}) to factorize $\det(K)$ as
    \be
    \det(K)=f_-(z_-,z_+,a,k)f_+(z_-,z_+,a,k),
    \label{det-K-factor}
    \ee
where
    \be
    f_\pm(z_-,z_+,a,k):=\frac{u}{2k^2}\:\sin(2ak)+e^{\pm
    2iak}\left(\frac{v}{k}\pm i\right),~~~~~
    u:=z_-z_+,~~~~~v:=\frac{z_-+z_+}{2}.
    \label{f-pm}
    \ee
Therefore $\det(K)=0$ if and only if at least one of
$f_\pm(z_-,z_+,a,k)$ vanishes.

Let us abbreviate $f_-(z_-,z_+,a,k)$ as $f(k)$, i.e., set
    \be
    f(k):=\frac{u}{2k^2}\:\sin(2ak)+e^{-
    2iak}\left(\frac{v}{k}- i\right).
    \label{f=}
    \ee
Then it is easy to see that $f_+(z_-,z_+,a,k)=-f(-k)$. Therefore,
the positive zeros of $f_+(z_-,z_+,a,k)$ are identical with the
absolute-value of the negative zeros of $f_-(z_-,z_+,a,k)$. In other
words, the spectral singularities are given by positive and negative
real zeros of $f(k)$. Another interesting property of $f(k)$ is that
it satisfies
    \be
    f(k)=-ie^{-2iak}\left[1+w_-+w_++w_-w_+(1-e^{4aik})\right]=
    -ie^{-2iak}{\rT}_{22}(k),
    \label{f=s22}
    \ee
where ${\rT}_{22}(k)$ is the the entry of the matrix $\rT$ of
(\ref{S}) with the row and column labels 2. According to
(\ref{f=s22}) the spectral singularities are the non-vanishing real
zeros of ${\rT}_{22}(k)$. This establishes the equivalence of
(\ref{def-1}) and (\ref{def-2}) for the double-delta-function
potentials.

In order to characterize the real zeros of $f(k)$, we set the real
and imaginary parts of the right-hand side of (\ref{f=}) equal to
zero. This gives
    \bea
    \left(-1+\frac{\Re(u)}{2k^2}+\frac{\Im(v)}{k}\right)\sin(2ak)+
    \left(\frac{\Re(v)}{k}\right)\cos(2ak)&=&0,
    \label{re-f=zero}\\
    \left(\frac{\Im(u)}{2k^2}-\frac{\Re(v)}{k}\right)\sin(2ak)+
    \left(\frac{\Im(v)}{k}- 1\right)\cos(2ak)&=&0.
    \label{im-f=zero}
    \eea
Because $\sin(2ak)$ and $\cos(2ak)$ cannot vanish simultaneously,
these equations hold provided that the matrix of coefficients of
$\sin(2ak)$ and $\cos(2ak)$ is singular. Equating the determinant of
this matrix to zero and simplifying the resulting equation, we find
    \be
    g(k):= k^3-2\Im(v)k^2+ \left(-\frac{\Re(u)}{2}+|v|^2\right)k+
    \frac{1}{2}\left[\Re(u)\Im(v)-\Re(v)\Im(u)\right]=0.
    \label{q21}
    \ee
Because $g$ is a real cubic polynomial, it always has at least one
real root $\kappa$. If $\kappa\neq 0$, $E_\star=\kappa^2$ is a
spectral singularity. Expressing $\kappa$ as a function of $u$ and
$v$ and inserting it in say (\ref{im-f=zero}) we find a sufficient
condition on the coupling constants $z_\pm$ for the existence of a
spectral singularity. Repeating this for all the roots of $g$ (for
the cases that (\ref{q21}) has other nonzero real solutions) we
obtain a complete characterization of the spectral singularities.
They lie on a three-dimensional surface ${\cal S}$ embedded in the
four-dimensional space $(\C^2)$ of the coupling constants
$(z_-,z_+)$. Figure~\ref{fig1} is a graphical demonstration of the
intersection of ${\cal S}$ with the plane $z_+=z_-^*$ that
represents the $\cP\cT$-symmetric region of $\C^2$. In the following
we examine some non-$\cP\cT$-symmetric regions of $\C^2$ and their
intersection with ${\cal S}$.
    \begin{enumerate}
    \item
    Consider the plane $\Pi_1$ in $\C^2$ defined by
    $z_+=-z_-^*=:z$ where $u=-|z|^2$ and $v=i\Im(z)$. Then
    Eqs.~(\ref{re-f=zero}) and (\ref{im-f=zero}) take the form
        \[\left[1+\left(\frac{\Re(z)}{k}\right)^2+\left(\frac{\Im(z)}
        {k}- 1\right)^2\right]\sin(2ak)=0=
        \left(\frac{\Im(z)}{k}- 1\right)\cos(2ak).\]
    These are satisfied if and only if $\sin(2ak)=0$ and $k=|\Im(z)|$.
    Therefore, we have a spectral singularity located at
    $E_\star=\kappa^2=\Im(z)^2$ if and only if
        \be
        \Im(z)=n\pi,~~~{\rm for~some}~~~n\in \Z-\{0\}.
        \label{anti-PT}
        \ee
    This shows that $\Pi_1$ intersects $\cal S$ along equidistant
    lines parallel to the $\Re(z)$-axis in $\Pi_1$.

    \item Consider the case that both $z_+$ and $z_-$ are
    purely imaginary. This also defines a plane in $\C^2$ that we
    denote by $\Pi_2$. In this case, we can express $z_\pm$ as
        $$z_\pm=:\frac{iy_\pm}{a},$$
    where $y_\pm$ are nonzero real numbers.
    In terms of $y_\pm$, Eqs.~(\ref{re-f=zero}) and (\ref{im-f=zero})
    take the form
        \[\left(\frac{y_+ +y_-}{2ak}-1 \right) \cos(2ak)=0=
        \left(\frac{y_+ +y_-}{2ak}-\frac{y_+ y_-}{2a^2k^2}-1 \right)
        \sin(2ak).\]
    There are two ways to satisfy these equations. Either
        \be
        \frac{y_+ +y_-}{2ak}-1=\sin(2ak)=0,
        \label{2-1}
        \ee
    or
        \be
        2a^2k^2-ak\left(y_+ +y_-\right)+y_+ y_-
        =\cos(2ak)=0.
        \label{2-2}
        \ee
    We consider these two cases separately.

    If (\ref{2-1}) holds, $E_\star=\kappa^2$ with
    $\kappa:=(y_++y_-)/2a$ is a spectral singularity provided
    that
            \be
            y_++y_-=\frac{n\pi}{2},~~~~n\in\Z-\{0\}.
            \label{pure-im-1}
            \ee
    This defines a set of equidistance parallel lines
    in $\Pi_2$ along which we have spectral singularities.

    If (\ref{2-2}) holds, $k=\kappa_n$ where $\kappa_n:=(2n+1)\pi/(4a)$
    for all $n\in \Z$, and
        \be
        y_+=a\kappa_n\left( \frac{ y_- -2a\kappa_n}{y_- -a\kappa_n}\right)=
        \frac{(2n+1)\pi}{2}\left[\frac{2y_--(2n+1)\pi}{
        4y_--(2n+1)\pi}\right].
        \label{pure-im-2}
        \ee
    This equation gives the location of another set of spectral
    singularities, namely $E_\star=\kappa_n^2$, in the plane $\Pi_2$.

    Figure~\ref{fig2} shows the curves in $\Pi_2$ along which a
    spectral singularity arises, i.e., $\Pi_2\cap\cS$.
        \begin{figure}[t]
        \begin{center}
        \vspace{-.5cm} \includegraphics[scale=.75,clip]{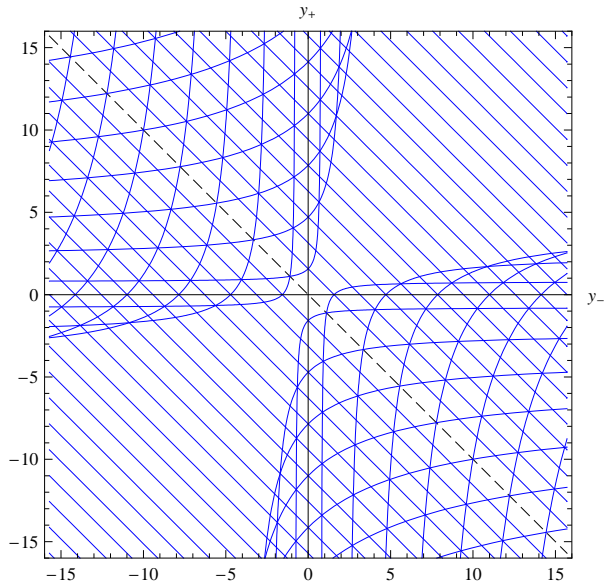}
    \parbox{14cm}{\caption{Curves in the $y_+$-$y_-$ plane
    ($\Pi_2$) along which one has a spectral singularity for
    purely imaginary couplings. There are spectral singularities
    along the $y_-$- and $y_+$-axes. The dashed line ($y_+=-y_-$)
    represents the $\cP\cT$-symmetric double-delta function
    potential with purely imaginary couplings. The intersection of
    these lines with the full curves correspond to the spectral
    singularities given by Eq.~(\ref{ss-imaginary-pt}).
    \label{fig2}}}\end{center}
    \end{figure}

    \item Consider the plane $\Pi_3$ in $\C^2$ corresponding
    to $z_+=-z_-=:z$. Then $v=0$ and $u=-z^2$. In particular,
    $\Im(u)=-2\Re(z)\Im(z)$. We can confine our attention to the
    subcase: $\Im(u)\neq 0$, because for $\Im(u)=0$ either $\Re(z)=0$,
    in which case $z_\pm$ are purely imaginary and the results of
    case~2 apply, or $\Im(z)=0$, in which case the potential is real
    and there are no spectral singularities.

    In view of $v=0$ and (\ref{q21}), $g(k)=k(k^2-\Re(u)/2)$.
    Therefore, $k$ does not have a real zero and there is no
    spectral singularities, if $\Re(u)\leq 0$. For $\Re(u)>0$, there
    is a spectral singularity at $E_\star=\kappa_\pm^2=\Re(u)/2$,
    where
        \be
        \kappa_\pm:=\pm \sqrt{\frac{\Re(u)}{2}}.
        \label{p3}
        \ee
    Inserting this equation in (\ref{im-f=zero}) gives
        \be
        \Im(u)=\pm\,\Re(u)\cot\left(a\sqrt{2\Re(u)}\right).
        \label{5}
        \ee
    Introducing the parameter $t:=a\sqrt{2\Re(u)}$,
    we can use (\ref{5}) to obtain the following parametric
    equations for the $r:=a\Re(z)$ and $s:=a\Im(z)$ values
    that correspond to the spectral singularities.
        \be
        |r(t)|=\frac{t}{2}\sqrt{|\csc t|-1},~~~~~
        |s(t)|=\frac{t\cos t}{\sqrt{|\sin t|-\sin^2t}}.
        \label{r-s=}
        \ee
    Figure~\ref{fig3} shows the graph of the parametric curves defined by
    (\ref{r-s=}). They form the intersection of the
    plane $\Pi_3$ with the singular region $\cS$ of $\C^2$.
\begin{figure}[t]
\vspace{-.5cm}
\begin{center}\includegraphics[scale=.75,clip]{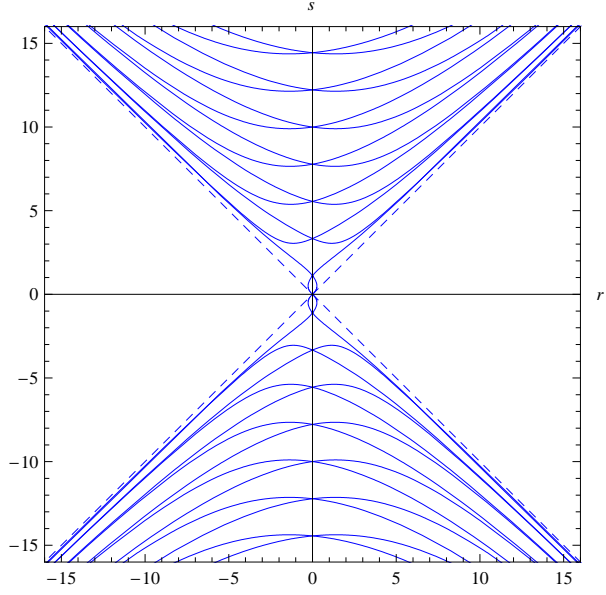}
\parbox{14cm}{\caption{Curves in the $r$-$s$ plane along which
spectral singularities occur for the coupling constants with
opposite sign. The origin $(s=r=0)$ does not actually lie on these
curves. The intersection of the curves with the $s$-axis corresponds
to the spectral singularities given by Eq.~(\ref{ss-imaginary-pt}).
The dashed lines are the lines $s=\pm r$.\label{fig3}}}\end{center}
\end{figure}

    \item Consider the case that $z_\pm=(1+is_\pm)/a$ with
    $s_\pm\in\R$ arbitrary. This corresponds to another plane in
    $\C^2$ that we denote by $\Pi_4$. Introducing
        \be
        s:=\frac{s_-+s_+}{2},~~~~~~t:=\frac{1+s_-s_+}{2},
        \label{case4-1}
        \ee
    we have
        \be
        v=\frac{1+is}{a},~~~~~u=\frac{1-t+is}{a^2}.
        \label{case4-2}
        \ee
    Inserting these in (\ref{q21}) yields
        \be
        a^3g(k)=(ak-s)(a^2k^2-ask+t)=0.
        \label{case4-3}
        \ee
    Therefore, we need to consider the following two possibilities.
    \begin{itemize}
    \item[]4.a) $k=s/a=(s_-+s_+)/(2a)$. In this case (\ref{im-f=zero}) is
    satisfied automatically while (\ref{re-f=zero}) yields
        \be
        t=s\cot(2s)+1.
        \label{case4-4}
        \ee
    We can use (\ref{case4-1}) and (\ref{case4-4})
    to express $s_\pm$ in terms of $s$. This gives
        \bea
        s_-&=&s\mp\sqrt{s^2+1-2(s\cot(2s)+1)},
        \label{case4-4m}\\
        s_+&=&2s-s_-=s\pm\sqrt{s^2+1-2(s\cot(2s)+1)}.
        \label{case4-4p}
        \eea

    \item[]4.b) $k\neq s/a$. Then according to (\ref{case4-3}),
        \be
        t=ask-a^2k^2.
        \label{case4-5}
        \ee
    Furthermore, both (\ref{re-f=zero}) and (\ref{im-f=zero}) become
        \be
        \tan(2ak)+ak=0
        \label{case4-6}
        \ee
    This equation has a countably infinite set of real solutions
    $\kappa_n$ that can be easily obtained numerically. Substituting
    $\kappa_n$ for $k$ in (\ref{case4-5}) and using (\ref{case4-1}),
    we find
        \bea
        s_-&=&s\mp\sqrt{s^2+1-2(as\kappa_n-a^2\kappa_n^2)},
        \label{case4-7b}\\
        s_+&=&2s-s_-=s\pm\sqrt{s^2+1-2(as\kappa_n-a^2\kappa_n^2)}.
        \label{case4-7a}
        \eea
    \end{itemize}
\begin{figure}[t]
\vspace{-.5cm}
\begin{center}\includegraphics[scale=.75,clip]{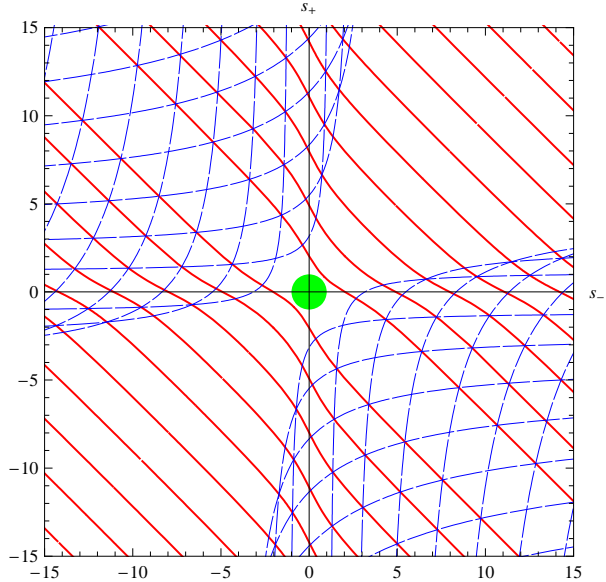}
\parbox{14cm}{\caption{Curves in the $s_-$-$s_+$ plane along which
the spectral singularities occur for the coupling constants of the
form $z_\pm=1+is_\pm$. The solid (red) and dashed (blue) curves
correspond to the spectral singularities with $k=(s_++s_-)/2$ (Case
4.a) and $k\neq(s_++s_-)/2$ (Case 4.b), respectively. Also shown (in
green) is the unit disc: $s_-^2+s_+^2\leq 1$, where there are no
spectral singularities. \label{fig4-new}}}\end{center}
\end{figure}
Figure~\ref{fig4-new} shows the parametric plot of the curves in the
$s_-$-$s_+$ plane corresponding to the spectral singularities for
both cases 4.a and 4.b with $a=1$. As seen from this figure, there
are no spectral singularity in the unit disc defined by
$s_-^2+s_+^2\leq 1$.

    \end{enumerate}

\subsection{Location of the Spectral Singularities and the Bound States}

As we noted in Section~2, the spectral singularities are given by
the real zeros of ${\rT}_{22}(k)$ while the bound states correspond
to zeroes of ${\rT}_{22}(k)$ with positive imaginary part. For the
double-delta-function potential, we can write ${\rT}_{22}(k)=0$ in
the following more compact form.
    \be
    (\fK-\fz_-)(\fK-\fz_+)=\fz_-\fz_+\:e^{2\fK},
    \label{e1}
    \ee
where we have used (\ref{S}) and introduced
    \[\fz_\pm:=a z_\pm=\frac{2m\alpha\zeta_\pm}{\hbar^2},
    ~~~~~~\fK:=2iak.\]
In particular, the spectral singularities are give by
    \be
    E_\star:=-\frac{\fK^2}{4a^2}
    \label{E=ss-bs}
    \ee
where $\fK$ is a nonzero solution of (\ref{e1}) lying on the
imaginary axis in the complex $\fK$-plane, i.e.,
    \[\fK\in\ell:=\{w\in\C~|~\Re(w)=0\neq w~\},\]
whereas the bound state ``energies'' are given by (\ref{E=ss-bs})
for solutions $\fK$ of (\ref{e1}) lying to the left of this axis,
i.e.,
    \[\fK\in\Pi_-:=\{w\in\C~|~\Re(w)< 0~\}.\]

\begin{figure}[pt]
    \begin{center} 
    \includegraphics[scale=.75,clip]{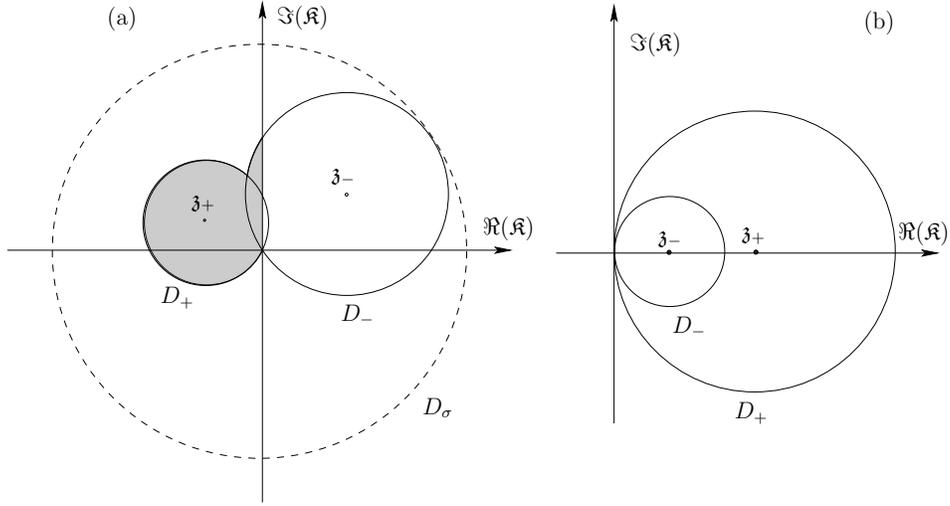}
    \parbox{15cm}{\caption{(a) Discs $D_\pm$ and $D_\sigma$
    for generic values of $\fz_\pm$. The gray area with the origin
    excluded is the region $\cR_{\vec\fz}$ where the bound states and
    spectral singularities are located (if any). (b) $D_\pm$
    for $\fz_\pm\in\R^+$. In this case $\cR_{\vec\fz}$ is empty.
    \label{fig4}}}\end{center}
    \end{figure}

For both spectral singularities and bound states, we have
$\Re(\fK)\leq 0$ which implies $|e^{2\fK}|\leq 1$. Taking the
modulus of both sides of (\ref{e1}), we find
    \be
    |\fK-\fz_-||\fK-\fz_+|\leq |\fz_+||\fz_-|.
    \label{e2}
    \ee
This is violated for any $\fK$ fulfilling
    \be
    |\fK-\fz_-|>|\fz_-|~~~{\rm and}~~~|\fK-\fz_+|>|\fz_+|.
    \label{e3}
    \ee
Therefore, the solutions of (\ref{e1}) with $\Re(\fK)\leq 0$ must
belong to the union of the discs
    \[D_{\pm}:=\big\{\fK\in\C~\big|~\big|\fK-\fz_\pm|
    \leq |\fz_\pm|~\big\}.\]
This provides an upper bound on the size of the region in the
complex $\fK$-plane where bound state energies and spectral
singularities are located, namely
    \[ \cR_{\vec\fz}:=(\Pi_-\cup\ell)\cap (D_+\cup D_-).\]
Here we have used the index $\vec\fz:=(\fz_-,\fz_+)$ to emphasize
the $\fz_\pm$-dependence of $\cR_{\vec\fz}$. Figure~\ref{fig4}
illustrates the discs $D_\pm$  and the region $\cR_{\vec\fz}$ for a
generic choice of $\fz_\pm$ and also for the case that $\fz_\pm$ are
real and positive. It is easy to see that in the latter case
$\cR_{\vec\fz}$ is empty and there are no spectral singularities or
bound states.

Let $D_\sigma$ and $\fD_\sigma$ be the disc and half-disc defined by
    \be
    D_\sigma:=\big\{\fK\in\C~\big|~|\fK|\leq\sigma~\big\},~~~~
    \fD_{\sigma}:=\big\{\fK\in\C~\big|~|\fK|\leq\sigma,~\Re(\fK)
    \leq 0.~\big\},
    \label{discs-z}
    \ee
where $\sigma$ is the largest of $2|\fz_\pm|$, i.e.,
    \be
    \sigma:= 2\:{\rm max}(|\fz_-|,|\fz_+|).
    \label{sigma=}
    \ee
Then, $\fK\in \fD_{\sigma}$ is a weaker necessary condition for the
existence of bound states and spectral singularities. This is simply
because $D_{\pm}\subseteq D_\sigma$. See Figure~\ref{fig4}(a).

Because the spectral singularities and bound states are given by the
zeros of
    \be
    F_{\vec\fz}(\fK):=(\fK-\fz_-)(\fK-\fz_+)-\fz_-\fz_+\:e^{2\fK},
    \label{F=}
    \ee
which is an entire (everywhere complex-analytic) function, and these
zeros are contained in $\fD_{\sigma}$ which is a compact subset of
the complex $\fK$-plane, we can determine the location of the
spectral singularities and bound states using the following
well-known result of complex analysis:
    \begin{itemize}
    \item[]\textbf{Theorem~1:} Let $C$ be a counterclockwise oriented
    contour bounding a compact and simply-connected region $R$ in
    complex plane and $h:\C\to\C$ be a function that is analytic on
    an open subset containing $R$. Then $h$ has a finite number of
    zeros in $R$. Moreover, if none of these zeros lie on $C$, the
    contour integral
        \be
        n_C:=\frac{1}{2\pi i}\oint_C\frac{h'(w)}{h(w)}\:dw
        \label{n=}
        \ee
    gives the sum of orders of zeros of $h$ contained in $R$.
    In particular, if all of these zeros are simple (of order 1),
    $n_C$ gives their number,
    \cite[\S10]{howie}.
    \end{itemize}

A proper use of this theorem requires a careful analysis of the
order of zeros of $F_{\vec\fz}$. It is not difficult to show that
the zeros of $F_{\vec\fz}$ can at most be of order three. Moreover,
$\fK$ is a third order zero of $F_{\vec\fz}$ if and only if $\fK=0$
and
    \be
    \fz_-=\frac{-1\pm i}{2},~~~~\fz_+=\frac{1}{2\fz_-}=
    \frac{-1\mp i}{2}.
    \label{3rd-order}
    \ee
This does not correspond to a spectral singularity or a bound state.
$F_{\vec\fz}$ has a second order zero $\fK_2$ if and only if
    \bea
    &&2\fz_-\fz_+\: e^{1+\fz_-+\fz_+\pm\sqrt{1+(\fz_--\fz_+)^2}}=
    1\pm\sqrt{1+(\fz_--\fz_+)^2},
    \label{2nd-1}\\
    &&\fK_2=\frac{1}{2}\left[1+\fz_-+\fz_+
    \pm\sqrt{1+(\fz_--\fz_+)^2}\right].
    \label{2nd-2}
    \eea
Requiring that $\Re(\fK_2)\leq 0$, we can use (\ref{2nd-1}) to show
that $|1\pm\sqrt{1+(\fz_--\fz_+)^2}|\leq 2|\fz_-\fz_+|$. Therefore,
there is no spectral singularity or bound state associated with a
second order zero of $F_{\vec\fz}$, if for both choices of the sign,
    \be
    \left|1\pm\sqrt{1+(\fz_--\fz_+)^2}\right|> 2|\fz_-\fz_+|.
    \label{2nd-3}
    \ee
This inequality in turn implies the following sufficient condition
for the lack of spectral singularities and bound states associated
with a second order zero of $F_{\vec\fz}$.
    \be
    |\fz_-\fz_+|(|\fz_-\fz_+|-1)<\frac{|\fz_--\fz_+|^2}{4}.
    \label{2nd-3n}
    \ee
In particular, such bound states or spectral singularities are
forbidden if $|\fz_-\fz_+|\leq 1$.

\begin{figure}[t]
    \begin{center}
    \includegraphics[scale=.90,clip]{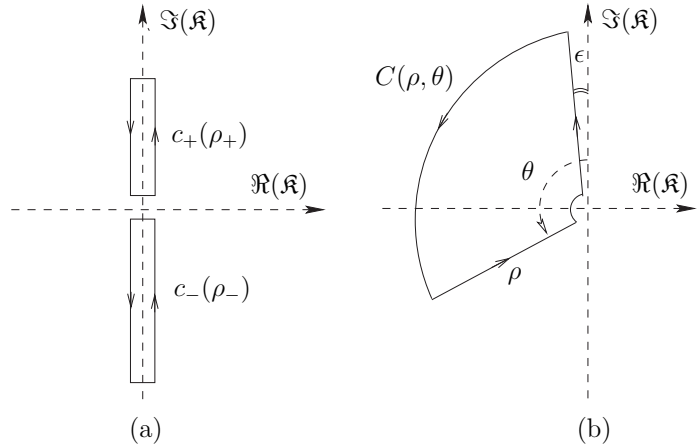}
    \parbox{14cm}{\caption{(a) $c_\pm(\rho_\pm)$ are rectangular contours
    of width $\epsilon\ll 1$ and hight $\rho_\pm\leq\sigma$; (b)
    $C(\rho,\theta)$ is the boundary of the region lying between
    circular arcs of side length $\epsilon\ll 1$ and
    $\rho\leq\sigma$ and opening angle
    $\theta\in[\epsilon,\pi-\epsilon]$.
    \label{fig5}}}\end{center}
    \end{figure}

Next, we return to the idea of using Theorem~1 for locating the
spectral singularities and bound states. For this purpose we can use
the contours $C(\rho,\theta)$ and $c_\pm(\rho_\pm)$ depicted in
Figure~\ref{fig5} to compute
    \bea
    n_\pm(\rho)&:=&\frac{1}{2\pi i}\oint_{c_\pm(\rho)}
    \frac{F_{\vec\fz}'(\fK)}{F_{\vec\fz}(\fK)}\:d\fK,
    \label{n-ss-pm}\\
    n(\rho_-,\rho_+)&:=&n_-(\rho_-)+n_+(\rho_+),
    \label{n-ss}\\
    N({\rho,\theta})&:=&\frac{1}{2\pi i}\oint_{C(\rho,\theta)}
    \frac{F_{\vec\fz}'(\fK)}{F_{\vec\fz}(\fK)}\:d\fK,
    \label{n=bs}
    \eea
where $\rho,\rho_\pm\in[\epsilon,\sigma]$, $\epsilon\ll 1$, and
$\theta\in[\epsilon,\pi-\epsilon]$. In the generic case where
$F_{\vec\fz}$ has no second order zeros, $n_\pm(\rho)$ and
$N({\rho,\theta})$ give the number of zeros of $F_{\vec\fz}$
enclosed by $c_\pm(\rho)$ and $C(\rho,\theta)$, respectively.
Therefore, plotting $n({\rho_-,\rho_+})$ and $N({\rho,\theta})$ as
functions of $\rho_\pm$ and $({\rho,\theta})$, we can locate all the
spectral singularities and bound states of the double-delta-function
potential for given coupling constants $z_\pm$. In particular, for
$\epsilon\to 0$, $n_{\rm tot}:=n(\sigma,\sigma)$ and $N_{\rm
tot}:=N(\sigma,\pi-\epsilon)$ respectively give the total number of
spectral singularities and bound states, except for the cases that
for some imaginary $\fK$ both $\fK$ and $-\fK$ are zeros of
$F_{\vec\fz}$. In the latter case, $\fK$ and $-\fK$ give rise to the
same spectral singularity, and one must account for the
corresponding double counting in $n_{\rm tot}$.

    \begin{figure}[t]
    \begin{center} 
    \includegraphics[scale=.25,clip]{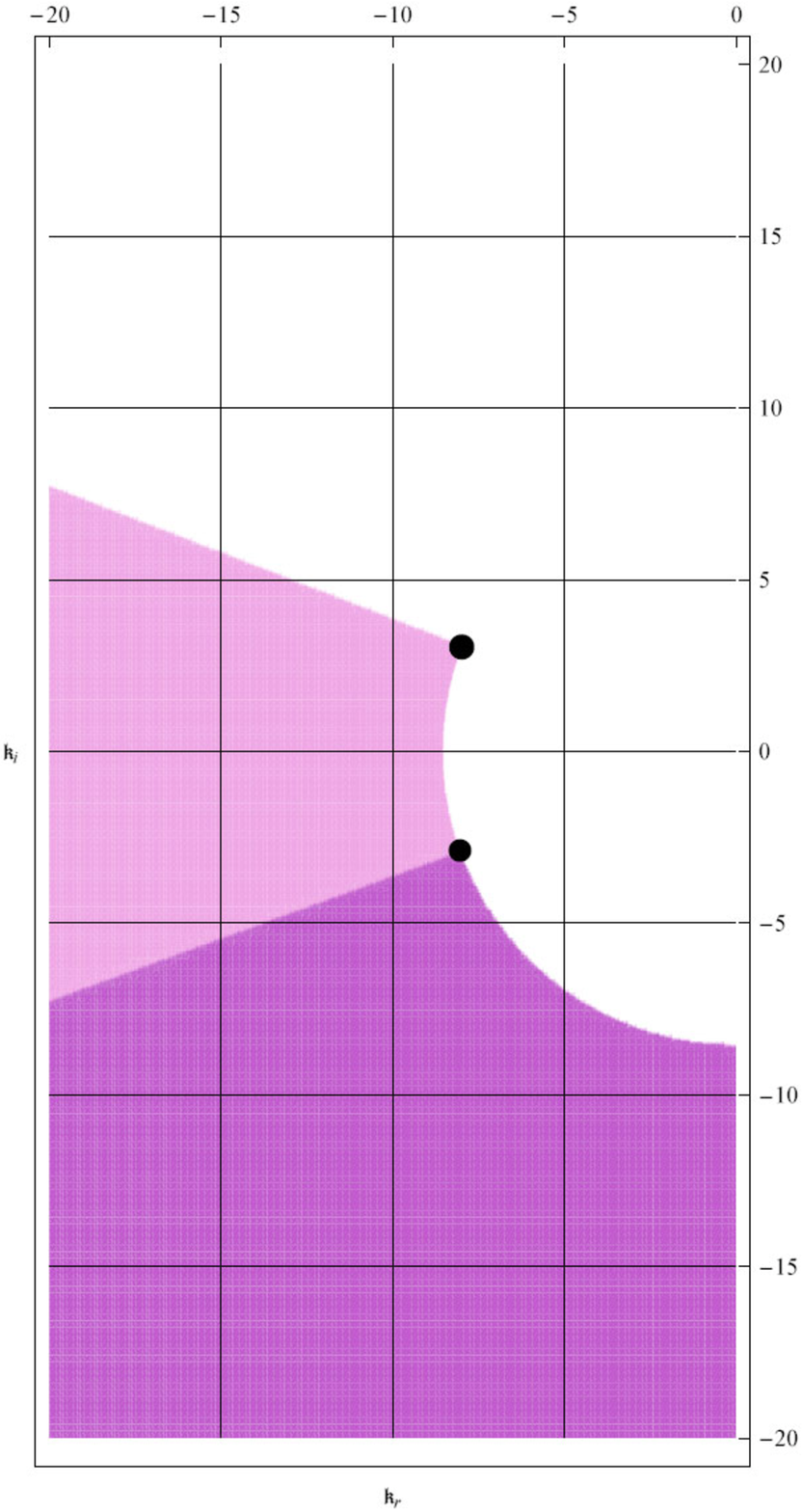}\hspace{1cm}
    \includegraphics[scale=.25,clip]{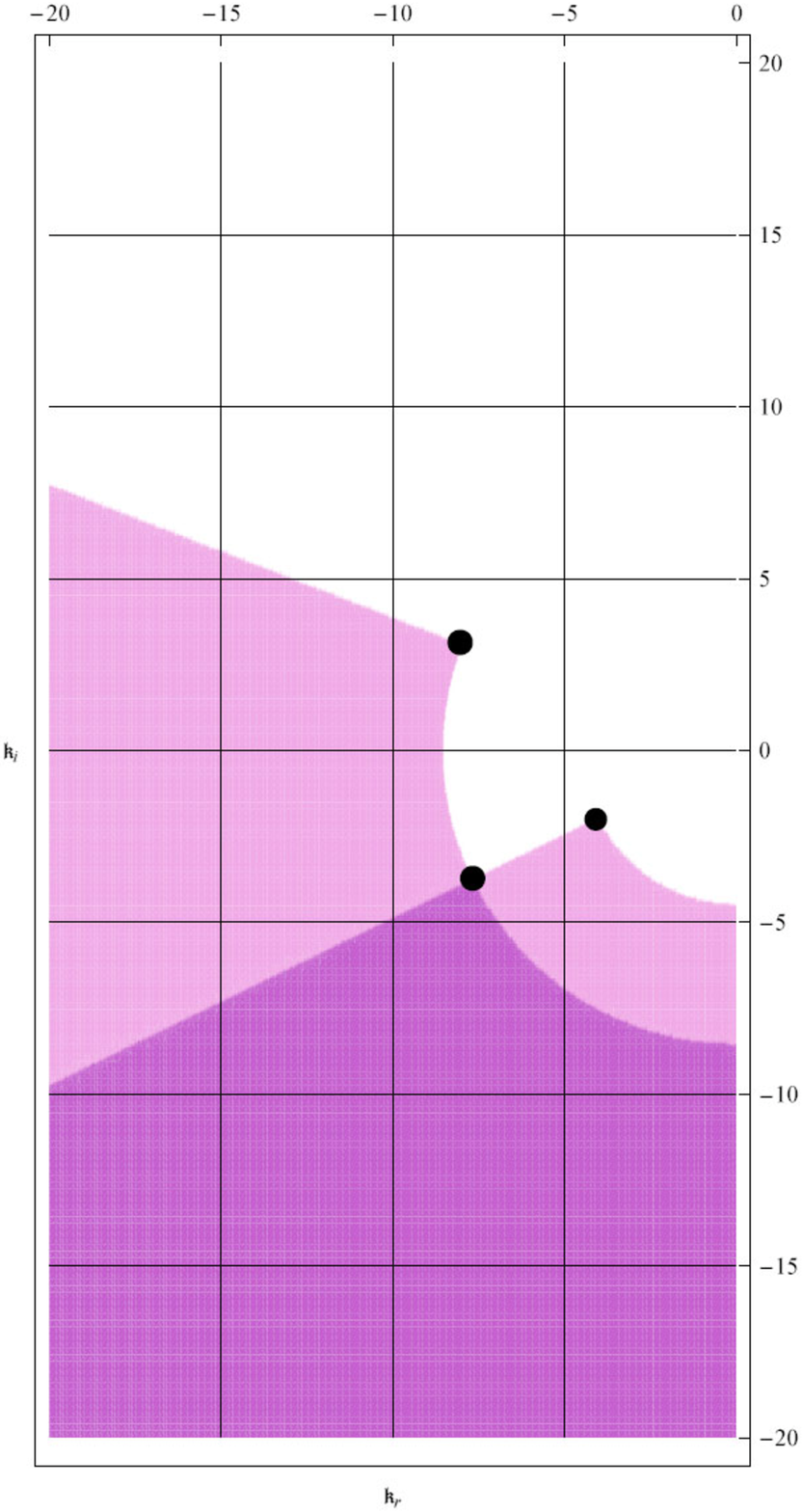}
    \parbox{15cm}{\caption{Density Plots of $N(\rho,\theta)$ for the
    $\cP\cT$-symmetric case $\fz_\pm=-8\pm 3i$ (on the left)
    and the non-${\cP\cT}$-symmetric case $\fz_-=-8 + 3 i$,
    $\fz_+=-4 -2 i$ (on the right) in the complex $\fK$-plane.
    $\fK_r$ and $\fK_i$ mark the real and imaginary axes.
    As the color changes from the lightest to the darkest
    $N(\rho,\theta)$ takes values 0,1, and 2, respectively.
    The critical points marked by black spots are the $\fK$-values
    corresponding to bound states. They are symmetric about the
    $\fK_r$-axis for the ${\cP\cT}$-symmetric case.
    \label{fig6}}}\end{center}
    \end{figure}
Note that locating spectral singularities is most conveniently
carried out using (\ref{q21}). In the absence of an analogous
equation giving the $k$ values for the bound states, we use
$N({\rho,\theta})$ to locate the latter. Figure~\ref{fig6} shows the
density plots of $N({\rho,\theta})$ for the $\cP\cT$-symmetric case
$\fz_\pm=-8\pm 3i$ and the non-${\cP\cT}$-symmetric case $\fz_-=-8 +
3 i$, $\fz_+=-4 - 2 i$. These resemble the phase diagrams of
statistical mechanics where the critical points correspond to the
bound states. As we expect, for the ${\cP\cT}$-symmetric case the
location of these points is symmetric about the real axis in the
complex $\fK$-plane.

    \begin{figure}[t]
    \begin{center} 
    \includegraphics[scale=1.4,clip]{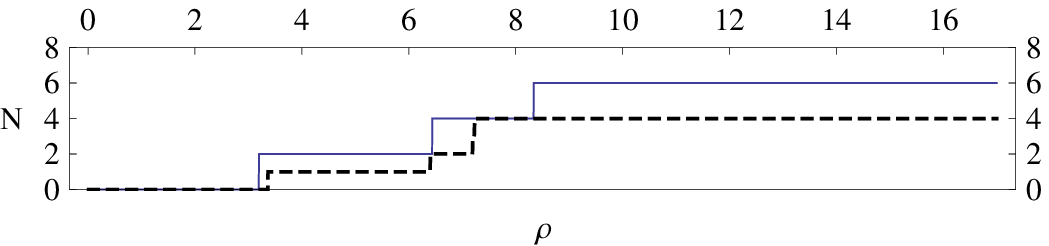}
    \vspace{1cm}

    \includegraphics[scale=1.4,clip]{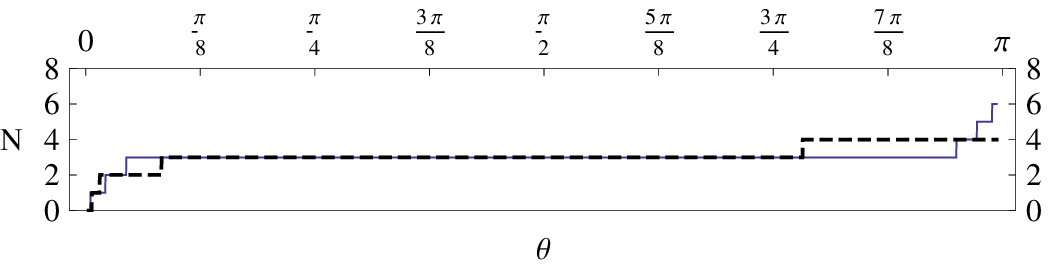}
    \parbox{15cm}{\caption{Plots of $N(\rho,\pi-.01)$ (top figure)
    and $N(\sigma,\theta)$ (bottom figure) for the
    $\cP\cT$-symmetric system defined by $\fz_\pm=-1\pm8i$ (the
    solid curves) and the non-$\cP\cT$-symmetric system defined by
    $\fz_-=-2 + 7 i$ and $\fz_+=-4 - 5 i$ (the dashed curves).
    For the $\cP\cT$-symmetric model $N(\rho,\pi-.01)$ changes
    in increments of 2 while $N(\sigma,\theta)$ is symmetric with
    respect to the $\theta=\pi/2$ line.
    \label{fig7}}}\end{center}
    \end{figure}
Figure~\ref{fig7} shows the graphs of $N(\rho,\pi-\epsilon)$ for the
$\cP\cT$-symmetric case $\fz_\pm=-1\pm 8i$ and the
non-${\cP\cT}$-symmetric case $\fz_-=-2 + 7 i$, $\fz_+=-4 - 5 i$.
These show the distance between the bound states from the origin.
For the $\cP\cT$-symmetric case the bound states are created in
complex-conjugate pairs with the same distance from the origin. This
explains the fact that the number of bound states changes in
increments of 2. This is clearly not the case for the
non-$\cP\cT$-symmetric case. For both of the above choices of the
coupling constants, $\sigma<17$. Therefore, the maximum value of
each curve gives the total number of bound states for the
corresponding system.

\begin{figure}[t]
    \begin{center}
    \includegraphics[scale=.40,clip]{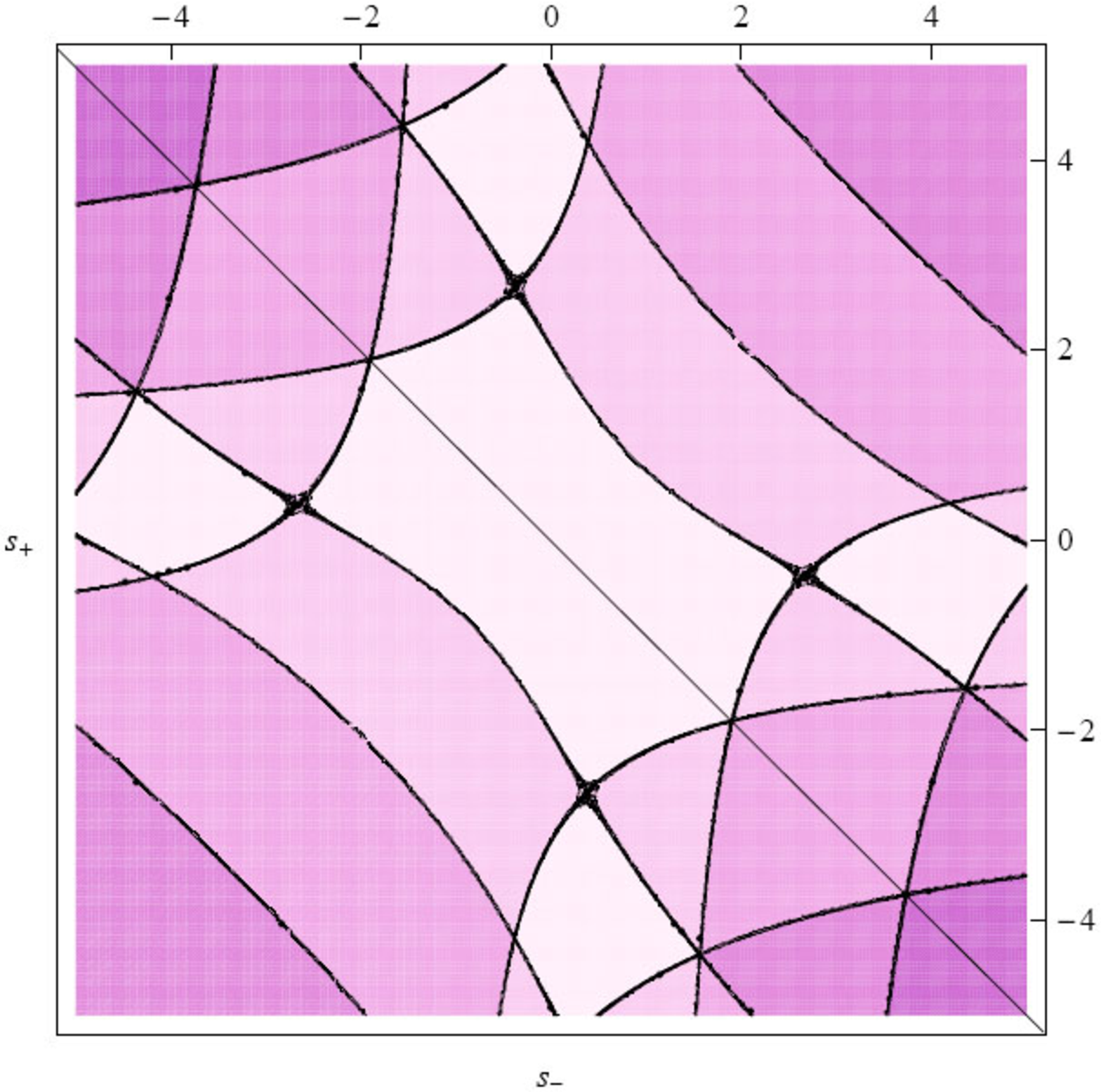}
    \parbox{15cm}{\caption{Contour plot of the number $N_{\rm tot}$
    of bound states located in the region: $\pi/2+\epsilon\leq
    {\rm arg}(\fK)\leq 3\pi/2-\epsilon$ for $\fz_\pm=1+s_\pm i$
    and $\epsilon=10^{-6}$. As the color changes from the lightest to
    the darkest $N_{\rm tot}$ take values $0,1,2,3,4$, respectively.
    The diagonal line $s_+=-s_-$ corresponds to the $\cP\cT$-symmetric
    region along which the number of bound states change in increments of
    2.
    \label{fig8}}}\end{center}
    \end{figure}
Figure~\ref{fig8} shows a contour plot of $N_{\rm
tot}:=N(\sigma,\pi-\epsilon)$ for $\fz_\pm=1+is_\pm$ and
$\epsilon=10^{-6}$ as functions of $s_\pm\in\R$. Although the real
part of the coupling constants are positive and equal, for large
enough values of their imaginary part the system develops bound
states. This is in contrast to the single delta-function potential
where there are no bound states for coupling constants with a
positive real part. We also see that in the $\cP\cT$-symmetric case
$s_+=-s_-$, which corresponds to the depicted diagonal line, the
number of bound states change in increments of 2. This is consistent
with the fact that these are produced in complex-conjugate pairs.

\subsection{Real Bound States and Quasi-Hermiticity}

An important feature of the graphical demonstration of the location
of spectral singularities and bound states in the complex
$\fK$-plane is that for the cases that $\Re(\fz_\pm)>0$ and
$|\Im(\fz_\pm)|$ are sufficiently small, the system does not have
any spectral singularities or bound states. Figures~\ref{fig4-new}
and~\ref{fig8} provide a clear demonstration of this phenomenon for
the case that $\Re(\fz_\pm)=1$.

The presence of spectral singularities is an obstruction to the
quasi-Hermiticity of the Hamiltonian operator. This is also true for
the bound states unless they happen to have real energies
(eigenvalues). We will refer to these bound states as ``\emph{real
bound states}.'' It is not difficult to see that generic bound
states are not real. In this subsection, we shall first derive
analytic expressions for the existence and location of real bound
states and then for fixed and positive values of $\Re(\fz_\pm)$ we
establish the existence of a positive lower bound on the size of a
region in the $\Im(\fz_-)$-$\Im(\fz_+)$ plane where the system is
free of both the spectral singularities and bound states. This is a
region where the Hamiltonian operator is quasi-Hermitian. It is in
this region that we can employ the machinery of pseudo-Hermitian
quantum mechanics \cite{jpa-2004b,review} to construct an associated
positive-definite metric operator and use the Hamiltonian operator
to define a unitary quantum system.

Excluding the case of a single-delta-function potential
\cite{jpa-2006b} where $\fz_+\fz_-=0$, we can express (\ref{e1}) as
    \be
    \left(\frac{\fK}{\fz_+}-1\right)\left(\frac{\fK}{\fz_-}-1\right)=
    e^{2\fK}.
    \label{e1-n}
    \ee
According to (\ref{E=ss-bs}), the real bound states are given by
real and negative solutions of (\ref{e1-n}). For these solutions the
right-hand side of (\ref{e1-n}) is real, positive, and less than 1.
Equating the left-hand side with its complex-conjugate yields
    \be
    \Im(\fz_-\fz_+)\fK=|\fz_-|^2\Im(\fz_+)+|\fz_+|^2\Im(\fz_-).
    \label{r-bs1}
    \ee
In order to explore the consequences of this equation we introduce
the notation
    \be
    r_\pm:=\Re(\fz_\pm),~~~~~~~~~s_\pm:=\Im(\fz_\pm),
    \label{notation}
    \ee
and consider the following cases separately.
    \begin{itemize}
    \item[(i)] $\Im(\fz_-\fz_+)=0$: In this case,
        \be
        r_-s_++r_+s_-=0,~~~~~~~
        \frac{s_-}{|\fz_-|^2}+\frac{s_+}{|\fz_+|^2}=0.
        \label{r-bs2}
        \ee
Therefore, either both $s_\pm$ vanish and the potential is real or
both $s_\pm$ are nonzero. In the latter case, (\ref{r-bs2}) implies
$\fz_-=\fz_+^*$. This is the $\cP\cT$-symmetric case for which
(\ref{e1}) reduces to
    \be
    |\fK-\fz_+|=|\fz_+|\,e^{\fK}.
    \label{e1-nn}
    \ee
Because $e^{\fK}<1$, this equation cannot be satisfied, if
$\Re(\fz_+)\geq 0$. This is consistent with the results of
\cite{demiralp}. Furthermore, for the non-${\cP\cT}$ cases with real
$\fz_-\fz_+$, such as $\fz_-=-\fz_+^*$ or imaginary $\fz_\pm$ with
$\fz_-\neq\fz_+^*$, there is no real bound states.

    \item[(ii)] $\Im(\fz_-\fz_+)\neq 0$: In this case, we can write
    (\ref{r-bs1}) as
        \be
        \fK=\frac{|\fz_-|^2\Im(\fz_+)+|\fz_+|^2\Im(\fz_-)}{\Im(\fz_-\fz_+)}.
        \label{r-bs3}
        \ee
    Substituting this equation in (\ref{e1}) gives a rather
    complicated relation between $\fz_-$ and $\fz_+$. This relation
    together with the requirement that the right-hand side of
    (\ref{r-bs3}) be negative provide the necessary and sufficient
    condition for the existence of real bound states for
    non-$\cP\cT$-symmetric cases. We have implemented this condition
    to address the existence of real bound states for the special
    cases where $\fz_+=\fz_-e^{i\nu}:=\fz\: e^{i\nu/2}$ with
    $\nu\in[0,2\pi)$. Figure~\ref{fig9} shows the curves in the
    complex $\fz$-plane along which real bound states exist for
    various values of $\nu$. It is important to note that all
    these curves are finite in length. Therefore, there are no real
    bound states for sufficiently large values of $|\fz|$.
    \begin{figure}[t]
    \begin{center}
    \includegraphics[scale=.9,clip]{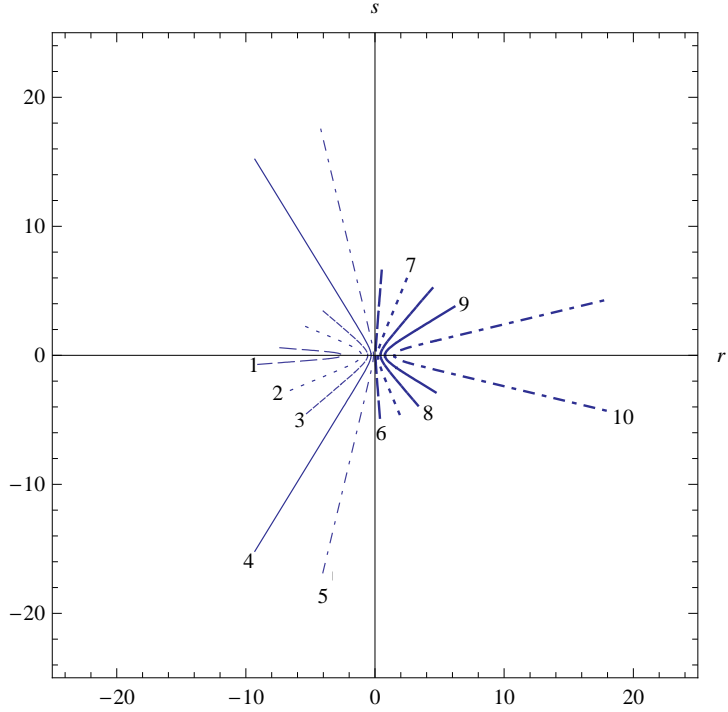}
    \parbox{13cm}{\caption{Curves in the complex $\fz$-plane along
    which real bound sates exist for $\fz_\pm=\fz e^{\pm i\nu/2}$,
    $\nu=\pi(2n-1)/20$, and $n\in\{1,2,\cdots,10\}$. The numbers
    attached to each curve segment is the corresponding value of $n$.
    $r$ and $s$ respectively mark the $\Re(\fz)$- and
    $\Im(\fz)$-axes. Note that all the curves have finite length.
    \label{fig9}}}\end{center}
    \end{figure}

    \end{itemize}

Next, we wish to show the existence of regions in the space of the
coupling constants $\fz_\pm$ where there is no spectral
singularities or bound states. Our main tools are the following
basic theorems of real and complex analysis.
    \begin{itemize}
    \item[] \textbf{Theorem~2:} Let $n\in\Z^+$, $D$ be a compact
    subset of $\R^n$ with its standard topology, and
    $\varphi:\R^n\to\R$ be a function that is continuous on $D$.
    Then $\{\varphi(\vec x)~\big|~\vec x\in D~\}$ has both
    a minimum and a maximum, \cite[\S 8]{Edwards}.
    \end{itemize}

    \begin{itemize}
    \item[] \textbf{Theorem~3 (Maximum Modulus Theorem):} Let $C$ be a
    contour bounding a compact and simply-connected subset $R$ of the complex
    plane and $h:\C\to\C$ be a function that is analytic on
    an open subset containing $R$. Then $\{|h(w)|~\big|~w\in R~\}$
    attains its maximum on $C$, \cite[\S III.1]{Lang}.
    \end{itemize}

First, we use Theorem~3 to prove the following preliminary results.
    \begin{itemize}
    \item[] \textbf{Lemma~1:} Let $\fD_\rho$ denote the following
    half disc of radius $\rho\in\R^+$:
        \[\fD_\rho:=\left\{~\fK\in\C~\big|~|\fK|\leq \rho,~\Re(\fK)\leq
        0~\right\},\]
    and $L:\C\to\C$ be the function defined by
        \be
        L(\fK):=\left\{
        \begin{array}{ccc}
        \frac{\mbox{\large$\displaystyle 1-e^{2\fK}$}}{
        \mbox{$\displaystyle\fK$}}
        &{\rm for}&\fK\neq 0,\\
        -2&{\rm for}&\fK= 0.\end{array}\right.
        \label{L=}
        \ee
    Then $|L|$ attains its maximum value on $\fD_\rho$ at
    $\fK=0$, i.e., $2=|L(0)|$ is the maximum of
    ${\cal A}_\rho:=\{|L(\fK)|~\big|~\fK\in \fD_\rho\}$ for all
    $\rho\in\R^+$.

    \item[] \textbf{Proof:} First, consider the case $\rho<1$. Then
    $\fD_\rho\subsetneq D_1$, which implies ${\cal A}_\rho\subseteq {\cal A}_1$.
    Therefore, the maximum of ${\cal A}_\rho$ is less than or equal
    to that of ${\cal A}_1$. This shows that it is sufficient to
    prove the lemma for the case $\rho\geq 1$. Because $L$ is an
    entire function and $\fD_\rho$ is compact, according to Theorem~3,
    ${\cal A}_\rho$ has a maximum that is located on the boundary of
    $\fD_\rho$. This is the union of the closed line segment
    $\ell_\rho:=\{ iy~|~y\in[-\rho,\rho]~\}$ and the open
    semicircle $C_\rho:=\{ i\rho~e^{i\varphi}~|~\varphi\in(0,\pi)~\}$.
    The maximum of ${\cal A}_\rho$ is the largest of the values taken by
    $|L|$ on $\ell_\rho$ and $C_\rho$. We will show that these values
    are bounded from above by $2$. Because $0\in \fD_\rho$ and $|L(0)|=2$,
    this is sufficient to prove the lemma. In the following we consider
    the values of $|L|$ on $\ell_\rho$ and $C_\rho$ separately.
        \begin{itemize}
        \item For all $\fK\in\ell_\rho$, we can write $\fK=iy$ for
        some $y\in[-\rho,\rho]$. Inserting $\fK=iy$ in (\ref{L=}) and
        computing the modulus of both sides of the resulting expression
        yields
            \be
            |L(\fK)|=\frac{2\sin y}{y}\leq 2.
            \label{bound-L1}
            \ee
        \item For all
        $\fK\in C_\rho$, we can write $\fK=i\rho~e^{i\varphi}$ for some
        $\varphi\in(0,\pi)$. Because $\rho\geq 1$ and
        $\sin\varphi>0$, (\ref{L=}) implies
            \be
            |L(\fK)|=\frac{\left|1-\exp(2i\rho~e^{i\varphi})\right|}{\rho}
            \leq 1+|\exp(2i\rho~e^{i\varphi})|=1+e^{-2\rho\sin\varphi}<2.
            \label{bound-L2}
            \ee
        This together with (\ref{bound-L1}) proves the lemma for
        $\rho\geq 1$. As we explained above this establishes the statement
        of the lemma also for the case $\rho<1$.~~~$\square$

         \end{itemize}

    \end{itemize}

    \begin{itemize}
    \item[] \textbf{Lemma~2:} Suppose that $r_\pm>0$. Then $\fK=0$
    is a first order zero of the function $F_{\vec\fz}$ defined by
    (\ref{F=}).
    \item[] \textbf{Proof:} Recall that the zeros of $F_{\vec\fz}$ are at
    most of order three and $F_{\vec\fz}(0)=0$. Therefore, it is
    sufficient to show that $\fK=0$ is not a second or third order
    zero of $F_{\vec\fz}$. Assume (by contradiction) that $\fK=0$ is
    a second order zero of $F_{\vec\fz}$. Then $F'_{\vec\fz}(0)=0$, i.e.,
    $\fz_-+\fz_-+2\fz_-\fz_+=0$. Equivalently, we have
        \[r_++r_-+2(r_-r_+-s_-s_+)=0,~~~~~s_++s_-+2(r_-s_++r_+s_-)=0.\]
    Solving the second of these for $s_-$ and inserting the result
    in the first, we find
        \[r_++r_-+2r_-r_++\frac{2(1+2r_-)s_+^2}{1+2r_+}=0.\]
    But this equation cannot be satisfied for $r_\pm>0$. This shows
    that the above assumption is false and $\fK=0$ is not a second
    order zero of $F_{\vec\fz}$. Next, we recall that $\fK=0$ is a
    third order zero of $F_{\vec\fz}$ if and only if (\ref{3rd-order})
    hold. But these conflict with the condition $r_\pm>0$.
    Hence $\fK=0$ is not a third order zero of $F_{\vec\fz}$.~~~$\square$
    \end{itemize}

Next, we use Theorem~2 and Lemmas~1 and~2 to prove the following
desired result.
    \begin{itemize}
    \item[] \textbf{Theorem~4:} Suppose that $r_\pm>0$ and
    $|s_\pm|<r_{\rm max}:={\rm max}(r_-,r_+)$. Then there is a positive
    upper bound $B_{\vec r}$ on $|s_\pm|$ such that for all $s_\pm$
    satisfying $|s_\pm|<B_{\vec r}$, the Hamiltonian (\ref{H}) does not
    have any spectral singularities or bound
    states.\footnote{Here and in what follows $\vec r:=(r_-,r_+)$.}

    \item[] \textbf{Proof:} Recall that spectral
    singularities and bound states are zeros $\fK$ of $F_{\vec\fz}$
    with $\Re(\fK)\leq 0$ and that they belong to $D_\sigma$, where
    $\sigma:=2\,{\rm max}(|\fz_-|,|\fz_+|)$. The
    latter is a subset of the half-disc
        \[\fD:=\fD_{\sqrt 8\,r_{\rm max}}=\Big\{\fK\in\C~\Big|~|~
        \fK|\leq \sqrt8\: r_{\rm max},~~\Re(\fK)\leq 0\Big\},\]
    because in view of $r_\pm\leq r_{\rm max}$ and $|s_\pm|<r_{\rm
    max}$, we have $\sigma<\sqrt 8\:r_{\rm
    max}$.
    According to Lemma~2, $\fK=0$ is a first order zero of
    $F_{\vec\fz}$. This implies that the function
    $G_{\vec\fz}:\C\to\C$ defined by
        \be
        G_{\vec\fz}(\fK):=\left\{\begin{array}{ccc}
        \fK^{-1}F_{\vec\fz}(\fK) & {\rm for} & \fK\neq 0,\\
        F'_{\vec\fz}(0)& {\rm for} & \fK= 0,\end{array}\right.
        \label{G=}
        \ee
    is an entire function and $G_{\vec\fz}(0)\neq 0$. Furthermore,
    the spectral singularities and bound states of the
    Hamiltonian~(\ref{H}) correspond to the zeros $\fK_0$ of
    $G_{\vec\fz}$ lying in $\fD$. Another important
    observation is that $G_{\vec r}$ has no zeros $\fK$ with
    $\Re(\fK)\leq 0$, because if existed these zeros would have
    corresponded to the spectral singularities or bound states of the
    Hamiltonian~(\ref{H}) with real and positive coupling constants
    $(\fz_\pm\in \R^+)$. But as we argued above this Hamiltonian does not
    have any spectral singularities or bound states. This observation
    establishes the fact that
        \be
        G_{\vec r}(\fK)\neq 0,~~~~~\mbox{for all}~\fK\in \fD.
        \label{G-r}
        \ee

    Because $G_{\vec r}$ is an entire function, $|G_{\vec r}|$ is
    continuous on $\fD$ which is a compact subset
    of $\C=\R^2$. In view of Theorem~2, this implies that the set
    $\{|G_{\vec r}(\fK)|~\big|~\fK\in \fD~\}$ has a minimum
    $m_{\vec r}$, i.e., there is $\fK_{\rm min}\in \fD$ such
    $m_{\vec r}=|G_{\vec r}(\fK_{\rm min})|$. Because
    $\fK_0,\fK_{\rm min}\in \fD$ and (\ref{G-r}) holds, we have
        \be
        0<|G_{\vec r}(\fK_{\rm min})|=m_{\vec r}\leq |G_{\vec r}(\fK_0)|.
        \label{bound-G}
        \ee
    Next, we introduce $J_{\vec\fz}:\C\to\C$
    as the function defined by
        \be
        J_{\vec\fz}(\fK):=G_{\vec\fz}(\fK)-G_{\vec r}(\fK).
        \label{J=}
        \ee
    Because $G_{\vec\fz}(\fK_0)=0$, we have
        \be
        |J_{\vec\fz}(\fK_0)|=|G_{\vec r}(\fK_0)|.
        \label{bound1}
        \ee
    Furthermore, in view of (\ref{G=}), (\ref{J=}),
    (\ref{L=}), and the fact that $\fK_0\neq 0$,
        \be
        J_{\vec\fz}(\fK_0)=-i(s_-+s_+)+
        \big[-s_-s_++i(r_-s_++r_+s_-)\big]L(\fK_0).
        \label{J=2}
        \ee
    This implies
        \bea
        |J_{\vec\fz}(\fK_0)|&\leq &
        |s_-|+|s_+|+
        \Big(|s_-||s_+|+|r_-||s_+|+|r_+||s_-|\Big)|L(\fK_0)|\nn\\
        &\leq &2\left(3r_{\rm max}+1\right)s_{\rm max},
        \label{bound-J}
        \eea
where $s_{\rm max}:={\rm max}(|s_-|,|s_+|)$ and we have used the
triangular inequality, the condition $|s_\pm|\leq r_{\rm max}$, and
$|L(\fK_0)|\leq 2$ that follows from Lemma~1.

If we combine (\ref{bound-J}) with (\ref{bound1}) and
(\ref{bound-G}), we obtain
    \be
    0<\frac{m_{\vec r}}{2(3r_{\rm max}+1)}\leq s_{\rm max}.
    \label{the-bound}
    \ee
This inequality is  violated for the values of $s_\pm$ for which
    \be
    |s_\pm|<\frac{m_{\vec r}}{2(3r_{\rm max}+1)}=:B_{\vec r}.
    \label{violate-bound}
    \ee
Therefore, for the cases that $|s_\pm|<B_{\vec r}$ the existence of
$\fK_0$ leads to a contradiction; such a $\fK_0$ cannot exist; and
there are no spectral singularities or bound states.~~~$\square$

\end{itemize}

The upper bound $B_{\vec r}$ given in (\ref{violate-bound}) involves
the minimum $m_{\vec r}$ of $|G_{\vec r}|$ on the half-Disc $\fD$.
Because $G_{\vec r}$ is a nowhere-zero analytic function on $\fD$,
$1/G_{\vec r}$ is also analytic on $\fD$. Hence, according to
Theorem~3, $1/|G_{\vec r}|$ attains its maximum $M_{\vec r}$ on the
boundary of $\fD$. It is not difficult to see that $m_{\vec
r}=1/M_{\vec r}$. Therefore, in practice, for given values of
$r_\pm$, we can obtain $m_{\vec r}$ by exploring the values of
$|G_{\vec r}|$ on the boundary of $\fD$.

\begin{figure}[t]
    \begin{center} 
    \includegraphics[scale=.7,clip]{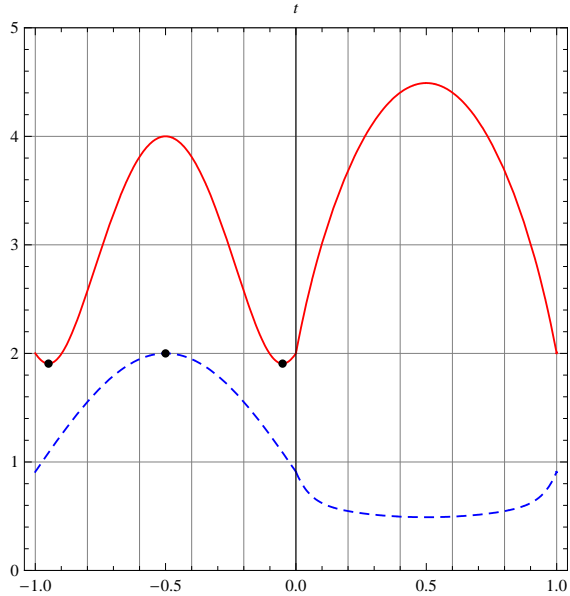}
    \parbox{15cm}{\caption{Plots of $|G_{\vec r}(\gamma_{\vec r}(t))|$
    (the full curve) and $|L(\gamma_{\vec r}(t))|$ (the dashed curve)
    as a function of $t\in[-1,1]$. $t_-\approx-0.949$ and $t_+\approx-.051$
    are the minimum points of $|G_{\vec r}(\gamma_{\vec r}(t))|$
    corresponding to $\fK_\pm\approx\pm 1.795 i$. These give the
    minimum value $m_{\vec r}\approx 1.906$. The maximum value of
    $|L(\gamma_{\vec r}(t))|$ is 2 that is attained at $t_0=-0.5$
    corresponding to $\fK=0$.
    \label{fig10}}}\end{center}
    \end{figure}

We can identify the boundary of $\fD$ with the graph $\Gamma_{\vec
r}$ of the parameterized curve:
    \be
    \gamma_{\vec r}(t):=2ir_{\rm
    max}\left[(2t+1)\Theta(-t)+e^{i\pi
    t}\Theta(t)\right],~~~~t\in[-1,1],
    \label{graph}
    \ee
where $\Theta$ is the unit step function: $\Theta(0):=1/2$ and
$\Theta(t):=(1+t/|t|)/2$ for $t\neq 0$. Figure~\ref{fig10} shows the
graphs of $|G_{\vec r}(\gamma_{\vec r}(t))|$ and $|L(\gamma_{\vec
r}(t))|$ for the case $r_\pm=1$ that is considered in
Figures~\ref{fig4-new} and~\ref{fig8}.
In this case, $r_{\rm max}=1$ and $\fD$ is the half-disc of radius
$2$ lying in $\Pi_-$. As seen from the graph of values of $|L|$, it
attains its maximum at $t=-0.5$ (corresponding to $\fK=0$) and its
maximum value is 2. This is consistent with the statement of
Lemma~1. The minimum points of $|G_{\vec r}|$ are located at
$t=-0.949$ and $t=-.051$. These correspond to $\fK_{\rm min}\approx
\pm 1.795 i$ where $|G_{\vec r}|$ takes its minimum value: $m_{\vec
r}\approx 1.906$. According to (\ref{violate-bound}), this gives
$B_{\vec r}=m_{\vec r}/8\approx 0.238$. Therefore, for $\fz_\pm=1\pm
is_\pm$ with $|s_\pm|<0.238$ there should be no spectral
singularities or bound states. This is in complete agreement with
the graphical data depicted in Figures~\ref{fig4-new} and
\ref{fig8}; the disc with center $s_\pm=0$ and radius 0.238 lies in
the region with no spectral singularities or bound states.

\section{Concluding Remarks}

In this article we provided an explicit demonstration of how
spectral singularities obstruct the existence of a biorthonormal
eigensystem and render the Hamiltonian non-diagonalizable. We
achieved this by obtaining a characterization of spectral
singularities in terms of the ${\rT}_{22}$ entry of the matrix $\rT$
of Eq.~(\ref{S-matrix}). In particular we showed that while bound
states are zeros of ${\rT}_{22}(k)$ with $\Im(k)>0$, the spectral
singularities are the real zeros of ${\rT}_{22}(k)$. It is not
difficult to infer from this observation that, similarly to the
bound states, the spectral singularities are linked with
singularities of the scattering matrix \cite{p89}. However, unlike
the bound states, they lie on the real axis in the complex
$k$-plane. This in turn suggests interpreting spectral singularities
as resonances having a vanishing width. Ref.~\cite{p89} provides a
thorough description of this interpretation and its physical
implications.

We established the utility of our general results by providing a
thorough analysis of the spectral properties of a two-parameter
family of complex point interactions. We obtained various results on
the nature and location of the bound states and spectral
singularities for this family and proved the existence of regions in
the space of coupling constants where both bound states and spectral
singularities are lacking and the Hamiltonian is quasi-Hermitian.

Throughout our study we examined the consequences of imposing
$\cP\cT$-symmetry which corresponds to restricting the coupling
constants to a complex plane in the space $\C^2$ of coupling
constants. This revealed a previously unnoticed fact that
$\cP\cT$-symmetric double-delta-function potential can involve
spectral singularities.

The results of this paper may be extended to complex point
interactions corresponding to three or larger number of delta
function potentials. Another line of research is to try to compute a
metric operator $\eta_{_{+}}$ and the corresponding equivalent
Hermitian Hamiltonian $h$ and the pseudo-Hermitian position and
momentum operators $X$ and $P$ for the double-delta-function
potential whenever the Hamiltonian is quasi-Hermitian. Theorem~4
provides the mathematical basis for a perturbative calculation of
$\eta_{_{+}}$, $h$, $X$, and $P$. We plan to report the results of
this calculation in a forthcoming publication.

\section*{Acknowledgments}

This work has been supported by the Scientific and Technological
Research Council of Turkey (T\"UB\.{I}TAK) in the framework of the
project no: 108T009, and by the Turkish Academy of Sciences
(T\"UBA). We wish to express our gratitude to Prof.\ Gusein Guseinov
for preparing and sending us a detailed description of spectral
singularities \cite{guseinov}.


\ed